\documentclass[preprint,review,12pt,sort&compress]{elsarticle}
\usepackage[margin=1.18 in]{geometry}
\usepackage{multirow}
\usepackage[usenames, dvipsnames]{color}
\definecolor{UW}{RGB}{64, 38, 96}

\usepackage{amssymb}

\usepackage{float}

\journal{Composite Structures}

\begin{document}
\begin{titlepage}

\clearpage\thispagestyle{empty}



\noindent

\hrulefill

\begin{figure}[h!]

\centering

\includegraphics[width=1.5 in]{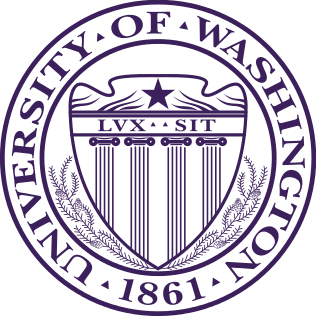}

\end{figure}


\begin{center}

{\color{UW}{

{\bf A\&A Program in Structures} \\ [0.1in]

William E. Boeing Department of Aeronautics and Astronautics \\ [0.1in]

University of Washington \\ [0.1in]

Seattle, Washington 98195, USA

}

}

\end{center} 

\hrulefill \\ \vskip 2mm

\vskip 0.5in

\begin{center}

{\large {\bf Study of the Fracturing Behavior of Thermoset Polymer Nanocomposites via Cohesive Zone Modeling}}\\[0.5in]

{\large {\sc Yao Qiao, Marco Salviato}}\\[0.75in]

{\sf \bf INTERNAL REPORT No. 18-07/02E}\\[0.75in]

\end{center}

\noindent {\footnotesize {{\em Submitted to Composite Structures \hfill July 2018} }}

\end{titlepage}

\newpage


\journal{Composite Structures}

\begin{frontmatter}


\cortext[cor1]{Corresponding Author, \ead{salviato@aa.washington.edu}}

\title{Study of the Fracturing Behavior of Thermoset Polymer Nanocomposites via Cohesive Zone Modeling}


\author[address]{Yao Qiao}
\author[address]{Marco Salviato\corref{cor1}}

\address[address]{William E. Boeing Department of Aeronautics and Astronautics, University of Washington, Seattle, Washington 98195, USA}

\begin{abstract}
\linespread{1}\selectfont

This work proposes an investigation of the fracturing behavior of polymer nanocomposites. Towards this end, the study leverages the analysis of a large bulk of fracture tests from the literature with the goal of critically investigating the effects of the nonlinear Fracture Process Zone (FPZ). 

It is shown that for most of the fracture tests the effects of the nonlinear FPZ are not negligible, leading to significant deviations from Linear Elastic Fracture Mechanics (LEFM) sometimes exceeding  $150$\% depending on the specimen size and nanofiller content. 

To get a deeper understanding of the characteristics of the FPZ, fracture tests on geometrically-scaled Single Edge Notch Bending (SENB) specimens are analyzed leveraging a cohesive zone model. It is found that the FPZ cannot be neglected and a bi-linear cohesive crack law generally provides the best match of experimental data.

\end{abstract}


\end{frontmatter}


\section{Introduction}
\label{intro}
The outstanding advances in polymer nanocomposites have paved the way for their broad use in engineering. Potential applications of these materials include microelectronics \cite{Rogers}, energy storage \cite{Yoo} and harvesting \cite{chih}, soft robotics \cite{zhaoxuanhe}, and bioengineering \cite{Joong}. One of the reasons of this success is that, along with remarkable enhancements of physical properties such as e.g. electric and thermal conductivity \cite{Ramirez,Balandin}, nanomodification offers significant improvements of stiffness \cite{jiang2013_1,Odegard}, strength \cite{konnola2015_1} and toughness \cite{zappalorto2013_1,zappalorto2013_2,CoryandYao,Hubert,Roy,Roy2}. These aspects make it an excellent technology to enhance the mechanical behavior of polymers \cite{zhang2008_1,johnsen2007_1,carolan2016_1,zamanian2013_1,dittanet2012_1,liu2011_1,WaJin13,ChaSei13,ChaSa14,kim2008_1,naous2006_1,wetzel2006_1,vaziri2011_1,newadd1} or to improve the weak matrix-dominated properties of fiber composites \cite{newadd2,pathak}.

While a large bulk of data on the mechanical properties of polymer nanocomposites is available already, an aspect often overlooked is the effect on the fracturing behavior of the region close to the crack tip featuring most of energy dissipation, the \textit{Fracture Process Zone} (FPZ). This is an important aspect since, due to the complex mesostructure characterizing nanocomposites, the size of the non-linear FPZ occurring in the presence of a large stress-free crack is usually not negligible \cite{Baz84,Baz90,bazant1996_1,bazant1998_1,salviato2016_1}. The stress field along the FPZ is nonuniform and decreases with crack opening, due to a number of damage mechanisms such as e.g. discontinuous cracking, micro-crack deflection, plastic yielding of nanovoids, shear banding and micro-crack pinning \cite{zhang2008_1,salviato2011_2, salviato2013_1,salviato2013_2,zappalorto2013_1,zappalorto2013_2,Schulte_14, quaresimin2014_1, Quaresimin_16,CoryandYao}. As a consequence, the fracturing behavior and, most importantly, the energetic size effect associated with the given structural geometry, cannot be described by means of classical Linear Elastic Fracture Mechanics (LEFM) which assumes the effects of the FPZ to be negligible. To seize the effects of a finite, non-negligible FPZ, the introduction of a characteristic (finite) length scale related to the fracture energy and the strength of the material is necessary \cite{Baz84,Baz90,bazant1998_1,bazant1996_1,salviato2016_1}.

This work proposes an investigation on the fracturing behavior of thermoset polymer nanocomposites with the goal of critically investigating the effects of the nonlinear Fracture Process Zone (FPZ). By employing Size Effect Law (SEL), a formulation endowed with a characteristic length inherently related to the FPZ size, and assuming a linear cohesive behavior \cite{cusatis2009_1}, a large bulk of literature data is analyzed. It is shown that for most of the fracture tests, the nonlinear behavior of the FPZ is not negligible, leading to significant deviations from LEFM. As the data indicate, this aspect needs to be taken into serious consideration since the use of LEFM to estimate mode I fracture energy can lead to an error as high as $157$\% depending on the specimen size and nanofiller content.

A cohesive zone model featuring a Linear Cohesive Law (LCL) is used to further understand the fracturing behavior of polymer nanocomposites. It is shown that while the LCL with corrected fracture energy by SEL is capable of capturing experimental data, this is not the case for the LCL with the fracture energy calculated by LEFM. This is the confirmation that Size Effect Law (SEL) can be adapted to re-analyze the fracture tests available in the literature for the first approximation. Taking advantage of size effect tests on thermoset-based graphene nanocomposites by Mefford \emph{et al.} \cite{CoryandYao}, it is also found that these materials are better described by a bi-linear cohesive law. As the results show, while the use of a linear cohesive law provides a good approximation, a bi-linear cohesive law provides a superior description of the fracturing behavior for different sizes.

\section{Quasi-brittle Fracture of Nanocomposites} 
\label{sec:literature}
\subsection{Size effect law for nanocomposites}
\label{sec:literatureapplication}
\begin{figure} [H]
\center
\includegraphics[scale=0.8]{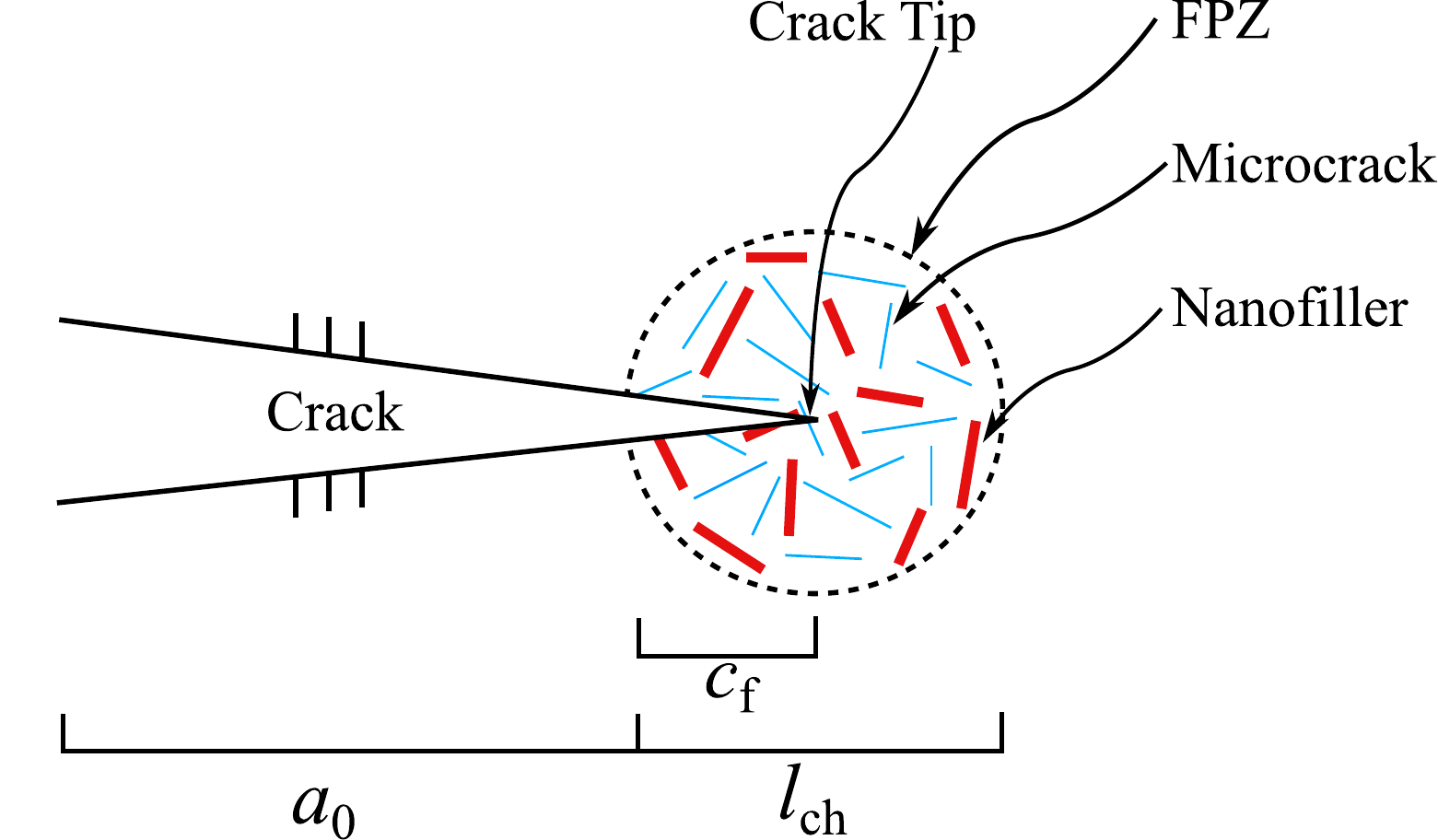}
\caption{Fracture Process Zone (FPZ) for thermoset polymer nanocomposites.}
\label{fig:FPZexample}
\end{figure}

\begin{figure} [H]
\center
\includegraphics[scale=0.5]{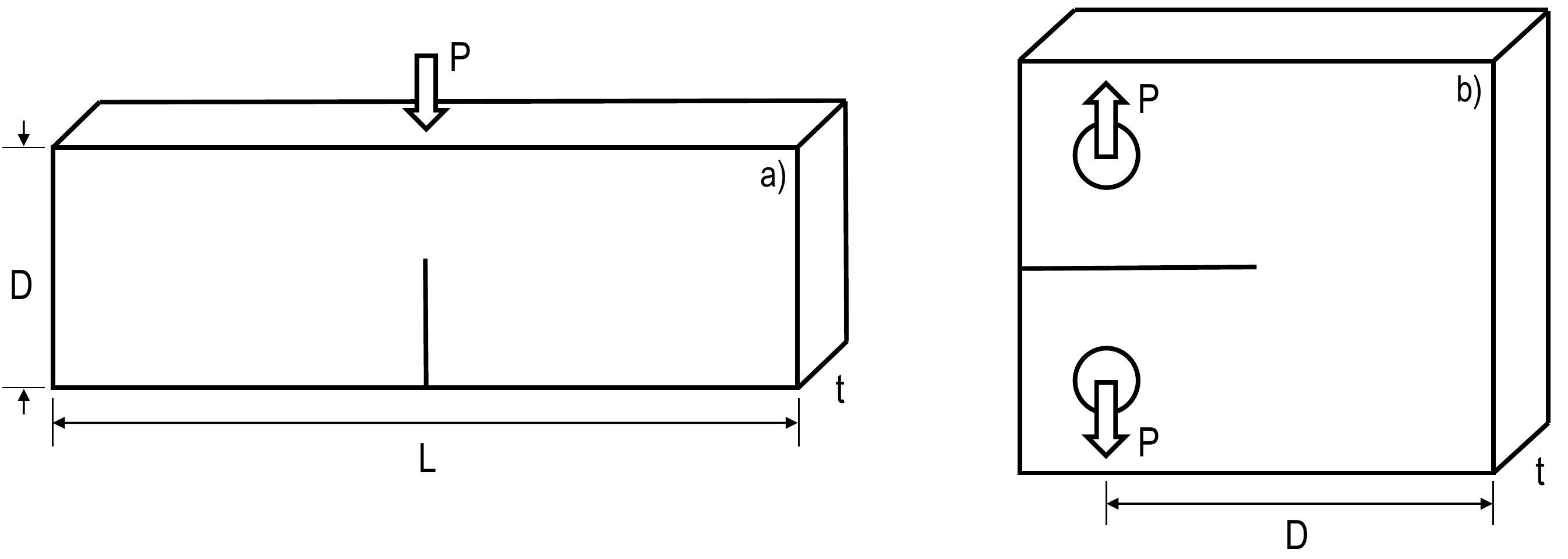}
\caption{Schematic representation of the SENB and CT specimens considered in this work.}
\label{fig:newgeometries}
\end{figure}

The fracture process in nanocomposites can be analyzed leveraging an equivalent linear elastic fracture mechanics approach to account for the presence of a FPZ of finite size as shown in Figure \ref{fig:FPZexample}. To this end, an effective crack length $a=a_0+c_f$ with $a_0=$ initial crack length and $c_f=$ effective FPZ length is considered. Following LEFM, the energy release rate can be written as follows:
\begin{equation}
G\left(\alpha\right)=\frac{\sigma_N^2D}{E^*}g(\alpha)
\label{eq:Gf}
\end{equation}
where $\alpha=a/D=$ normalized effective crack length, $E^*= E$ for plane stress and $E^*= E/\left(1-\nu^2\right)$ for plane strain, $g\left(\alpha\right)=$ dimensionless energy release rate and, $D$ is represented in Figure \ref{fig:newgeometries} for Single Edge Notch Bending (SENB) and Compact Tension (CT) specimens respectively. $\sigma_N$ represents the nominal stress defined as e.g. $\sigma_N=3PL/\left(2tD^2\right)$ for SENB specimens or $\sigma_N=P/\left(tD\right)$ for CT specimens where, following Figure \ref{fig:newgeometries}, $P$ is the applied load, $t$ is the thickness and $L$ is the span between the two supports for a SENB specimen as defined in ASTM D5045-99 \cite{ASTM_SENB}.

At incipient crack onset, the energy release rate ought to be equal to the fracture energy of the material. Accordingly, the failure condition can now be written as:
\begin{equation}
G\left(\alpha_0+c_f/D\right)=\frac{\sigma_{Nc}^2D}{E^*}g\left(\alpha_0+c_f/D\right)=G_f
\label{failure}
\end{equation}
where $G_f$ is the mode I fracture energy of the material and $c_f$ is the effective FPZ length, both assumed to be material properties. It should be remarked that this equation characterizes the peak load conditions if $g'(\alpha)>0$, i.e. only if the structure has positive geometry \cite{bazant1998_1}.

By approximating $g\left(\alpha\right)$ with its Taylor series expansion at $\alpha_0$ and retaining only up to the linear term of the expansion, one obtains:
\begin{equation}
G_f=\frac{\sigma_{Nc}^2D}{E^*} \left[g(\alpha_0)+\frac{c_f}{D}g'(\alpha_0)\right]
\label{Taylor}
\end{equation}
which can be rearranged as follows \cite{bazant1998_1}:
\begin{equation}
\sigma_{Nc}=\sqrt{\frac{E^*G_f}{Dg(\alpha_0)+c_fg'(\alpha_0)}}
\label{eq:Sel}
\end{equation}
where $g'\left(\alpha_0\right)=\mbox{d}g\left(\alpha_0\right)/\mbox{d}\alpha$.

This equation, known as Ba\v zant's Size Effect Law (SEL) \cite{Baz84,Baz90,bazant1998_1,salviato2016_1}, relates the nominal strength to mode I fracture energy, a characteristic size of the structure, $D$, and to a characteristic length of the material, $c_f$, and it can be rewritten in the following form:
\begin{equation}
\sigma_{Nc}=\frac{\sigma_{0}}{\sqrt{1+D/D_0}}
\label{eq:sigmaNc2}
\end{equation}
with $\sigma_0=\sqrt{E^*G_f/c_fg'(\alpha_0)}$ and $D_0=c_fg'(\alpha_0)/g(\alpha_0)=$ constant, depending on both FPZ size and specimen geometry. Contrary to classical LEFM, Eq. (\ref{eq:sigmaNc2}) is endowed with a characteristic length scale $D_0$. This is key to describe the transition from ductile to brittle behavior with increasing structure size.

\subsection{Calculation of $g\left(\alpha\right)$ and $g'\left(\alpha\right)$}
\label{sec:literatureforgeneralnanocomposites}

\subsubsection{Single Edge Notch Bending (SENB) specimens}
\label{sec:literatureSENB}
The calculation of $g(\alpha)$ and $g'(\alpha)$ for SENB specimens can be done according to the procedure described in \cite{CoryandYao}. This leads to the following polynomial expressions:
\begin{equation}
g(\alpha)=1155.4\alpha^5-1896.7\alpha^4+1238.2\alpha^3-383.04\alpha^2+58.55\alpha-3.0796
\end{equation}
\begin{equation}
g'(\alpha)=18909\alpha^5-31733\alpha^4+20788\alpha^3-6461.5\alpha^2+955.06\alpha-50.88
\end{equation}

\subsubsection{Compact Tension (CT) specimens}
\label{sec:literatureCT}
In the case of CT specimens, the values for $g(\alpha)$ and $g'(\alpha)$ can be determined leveraging the equations provided by ASTM D5045-99 \cite{ASTM_SENB}. Following the standard, the mode I Stress Intensity Factor (SIF), $K_I$, can be written as:
\begin{equation}
K_I=\frac{P}{t\sqrt{D}}f(\alpha)
\label{eq:K1cCT} \\
\end{equation}
where $\alpha=a/D$ and $D$ is the distance between the center of hole to the end of the specimen as defined in ASTM D5045-99 \cite{ASTM_SENB} (see Figure \ref{fig:newgeometries}b). The nominal stress $\sigma_N$ for CT specimens can be defined as:
\begin{equation}
\sigma_N=\frac{P}{tD}
\label{eq:stressnomCT}
\end{equation} 
The mode I Stress Intensity Factor can be rewritten as follows by combining Eq. (\ref{eq:K1cCT}) and Eq. (\ref{eq:stressnomCT}):
\begin{equation}
K_I=\sqrt{D}\sigma_Nf(\alpha)
\label{eq:K1cCT2}
\end{equation}
By considering the relationship between energy release rate and stress intensity factor for a plane strain condition, the mode I energy release rate results into the following expression:
\begin{equation}
G_I=\frac{D\sigma^2_{N}}{E}g(\alpha) 
\label{eq:gCT}
\end{equation} 
where $g(\alpha)=f^2(\alpha)(1-\upsilon^2)$, and $f(\alpha)$ is a dimensionless function accounting for geometrical effects and the finiteness of the structure (see e.g. \cite{ASTM_SENB}). Once $g(\alpha)$ is derived, the expression of $g'(\alpha)$ can be obtained by differentiation leading to the following polynomial expressions for $g(\alpha)$ and $g'(\alpha)$ respectively:
\begin{equation}
g(\alpha)=33325\alpha^5-52330\alpha^4+32016\alpha^3-9019.1\alpha^2+1230.1\alpha-51.944
\end{equation}
\begin{equation}
g'(\alpha)=555868\alpha^5-895197\alpha^4+554047\alpha^3-159153\alpha^2+21035\alpha-917.3
\end{equation}

\section{Fracture behavior of thermoset nanocomposites: analysis and discussion}
In the following sections, a large bulk of data on the fracturing behavior of nanocomposites is critically analyzed employing the expressions derived in Section \ref{sec:literature}. First, fracture tests data on the thermoset polymer reinforced by different nanoparticles are analyzed to investigate how the FPZ affects the failure behavior. Then, leveraging SEL and assuming a linear cohesive behavior, a large bulk of data from the literature originally elaborated by LEFM is re-analyzed to include the effects of the FPZ.

\subsection{Fracture Scaling of Nanocomposites} 
\label{sec:graphene}
\begin{figure} [H]
\center
\includegraphics[scale=0.7]{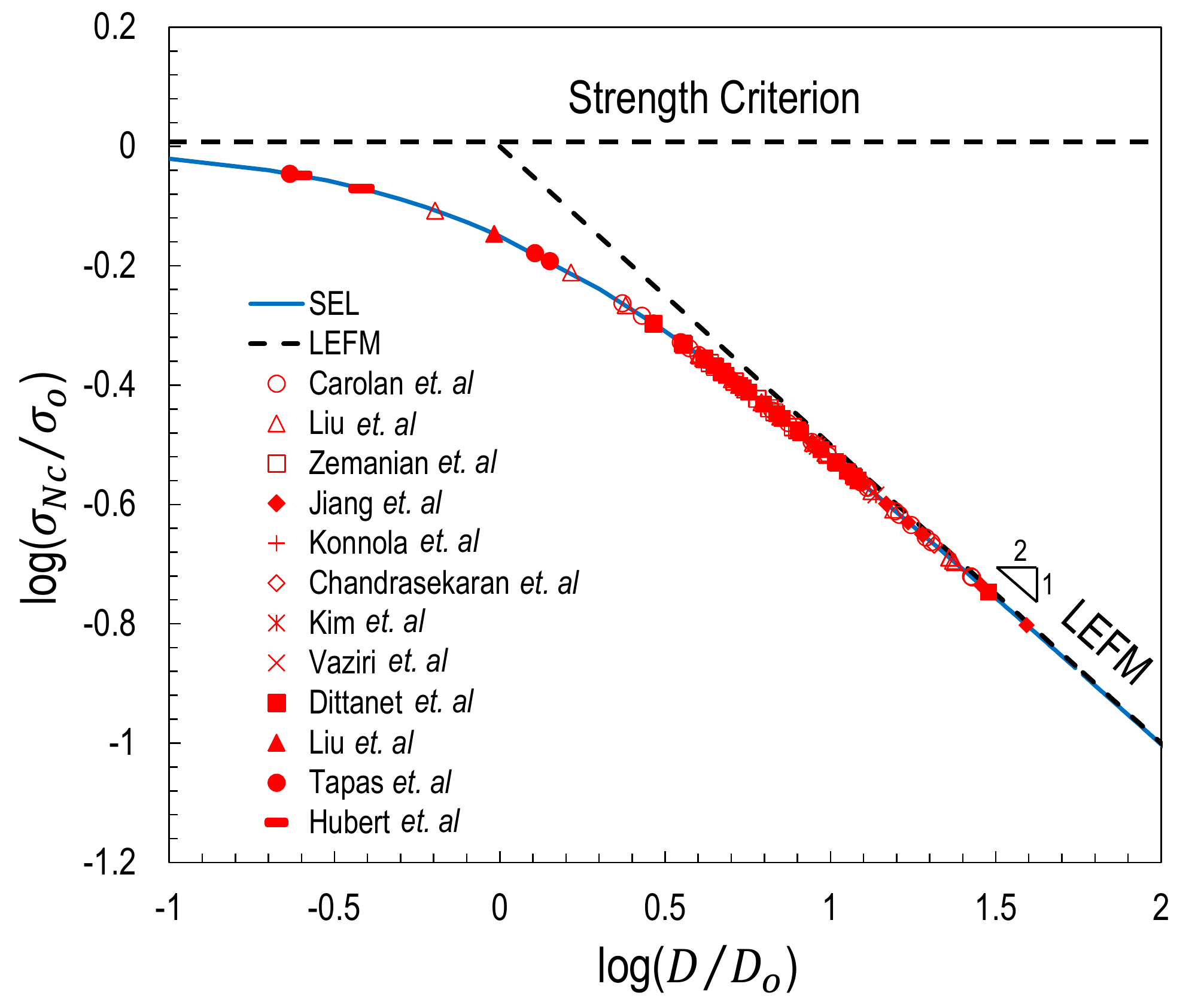}   
\caption{Size effect curves in polymer nanocomposites: data taken from the literature \cite{jiang2013_1,konnola2015_1,carolan2016_1,zamanian2013_1,ChaSa14,kim2008_1,vaziri2011_1,dittanet2012_1,liu2011_1,Tapas,Hubert2}.}
\label{fig:sizeeffectcurves}
\end{figure}

To investigate the effects of the non-linear FPZ, the fracture tests on the thermoset polymer reinforced by nanoparticles in the literature are analyzed and discussed. Figure \ref{fig:sizeeffectcurves} shows the normalized structural strength $\sigma_{Nc}/\sigma_{0}$ of the literature data as a function of the normalized structure size $D/D_0$ in double logarithmic scale. The solid line represents the fitting by SEL. In such a graph, the structural scaling predicted by LEFM is represented by a dashed line of slope $-1/2$ whereas the case of no scaling, as predicted by stress-based failure criteria, is represented by a horizontal line. The intersection between the LEFM asymptote, typical of brittle behavior, and the pseudo-plastic asymptote, typical of ductile behavior, corresponds to $D=D_0$, called the \emph{transitional size} \cite{bazant1998_1}. 

As can be noted from Figure \ref{fig:sizeeffectcurves}, the experimental data are in excellent agreement with SEL, which inherently captures the transition from strength-dominated to toughness-dominated fracture. More importantly, the figure shows that although some fracture tests reported in the literature were conducted under LEFM conditions (assumed by ASTM D5045-99 \cite{ASTM_SENB}), most of the data are located in the transitional region.

Accordingly, the experimental data show that LEFM does not always provide an accurate method to extrapolate the structural strength of larger structures from lab tests on small-scale specimens, especially if the size of the specimens belonged to the transitional zone. In fact, the use of LEFM in such cases may lead to a significant underestimation of structural strength, thus hindering the full exploitation of graphene nanocomposite fracture properties. This is a severe limitation in several engineering applications such as e.g. aerospace or aeronautics for which structural performance optimization is of utmost importance. On the other hand, LEFM always overestimates significantly the strength when used to predict the structural performance at smaller length-scales. This is a serious issue for the design of e.g. graphene-based MEMS and small electronic components or nanomodified carbon fiber composites in which the inter-fiber distance occupied by the resin is only a few micrometers and it is comparable to the FPZ size. In such cases, SEL or other material models characterized by a characteristic length scale ought to be used. 

\subsection{Effects of a finite FPZ on the calculation of Mode I fracture energy}
\label{sec:literatureforgeneralnanocomposites}
Notwithstanding the importance of understanding the scaling of the fracturing behavior, the tests conducted by Mefford \textit{et al}. \cite{CoryandYao} represent, to the best of the authors' knowledge, the only comprehensive investigation on the size effect in nanocomposites available to date. All the fracture tests reported in the literature were conducted on one size and analyzed by means of LEFM. Considering the remarkable effects of the nonlinear FPZ on the fracturing behavior documented in the foregoing section, it is interesting to critically re-analyze the fracture tests available in the literature by means of SEL. This formulation is endowed with a characteristic length related to the FPZ size and, different from LEFM, it has been shown to accurately capture the transition from brittle to quasi-ductile behavior of nanocomposites.

\subsubsection{Application of SEL to thermoset polymer nanocomposites}
\label{sec:literatureapplication}
To understand if the quasi-brittle behavior reported in previous tests \cite{CoryandYao} is a salient feature of graphene nanocomposites only or if it characterizes other nanocomposites, a large bulk of literature data are re-analyzed by SEL using Eq.(\ref{Taylor}) in order to study the effects of the FPZ. In this analysis, in the absence of data on the effective FPZ length, $c_f$, from the literature, it is assumed that $c_f=0.44 l_{ch}$ which, according to Cusatis \emph{et al.} \cite{cusatis2009_1}, corresponds to the assumption of a linear cohesive law. In this expression, $l_{ch}=E^*G_f/f_t^2$ is Irwin's characteristic length which depends on Young's modulus $E^*$, the mode I fracture energy $G_f$ and the ultimate strength of the material $f_t$. Substituting this expression into Eq. (\ref{Taylor}) and rearranging one gets the following expression which relates the fracture energy calculated according to SEL to the fracture energy calculated by LEFM:
\begin{equation}
G_{f,SEL}=\frac{G_{f,LEFM}}{1-\frac{0.44 E^*g'(\alpha_0)G_{f,LEFM}}{Df_t^2 g(\alpha_0)}}
\label{eq:Gf_SEL2}
\end{equation}
In this equation,  $G_{f,LEFM}=\sigma_{Nc}^2 Dg(\alpha_0)/E^*$ represents the fracture energy which can be estimated by analyzing the fracture tests by LEFM. 

It can be observed from Eq.(\ref{eq:Gf_SEL2}) that the correct fracture energy in the literature can be calculated by knowing three key parameters provided that $g(\alpha_0)$ and $g'(\alpha_0)$ are known: (1) the fracture energy through the use of LEFM; (2) the Young's modulus of the specimens at different nanoparticle concentrations; and (3) the ultimate strength of the specimens at different nanoparticle concentrations. For cases in which those parameters are not provided by the authors, the ultimate strength, Young's modulus, and Poisson's ratio of nanocomposites are reasonably assumed to be 50 MPa, 3000 MPa, and 0.35 respectively. 

\subsubsection{Mode I fracture energy of thermoset polymer nanocomposites}
\begin{figure}[H]
\center
\includegraphics[scale=0.5]{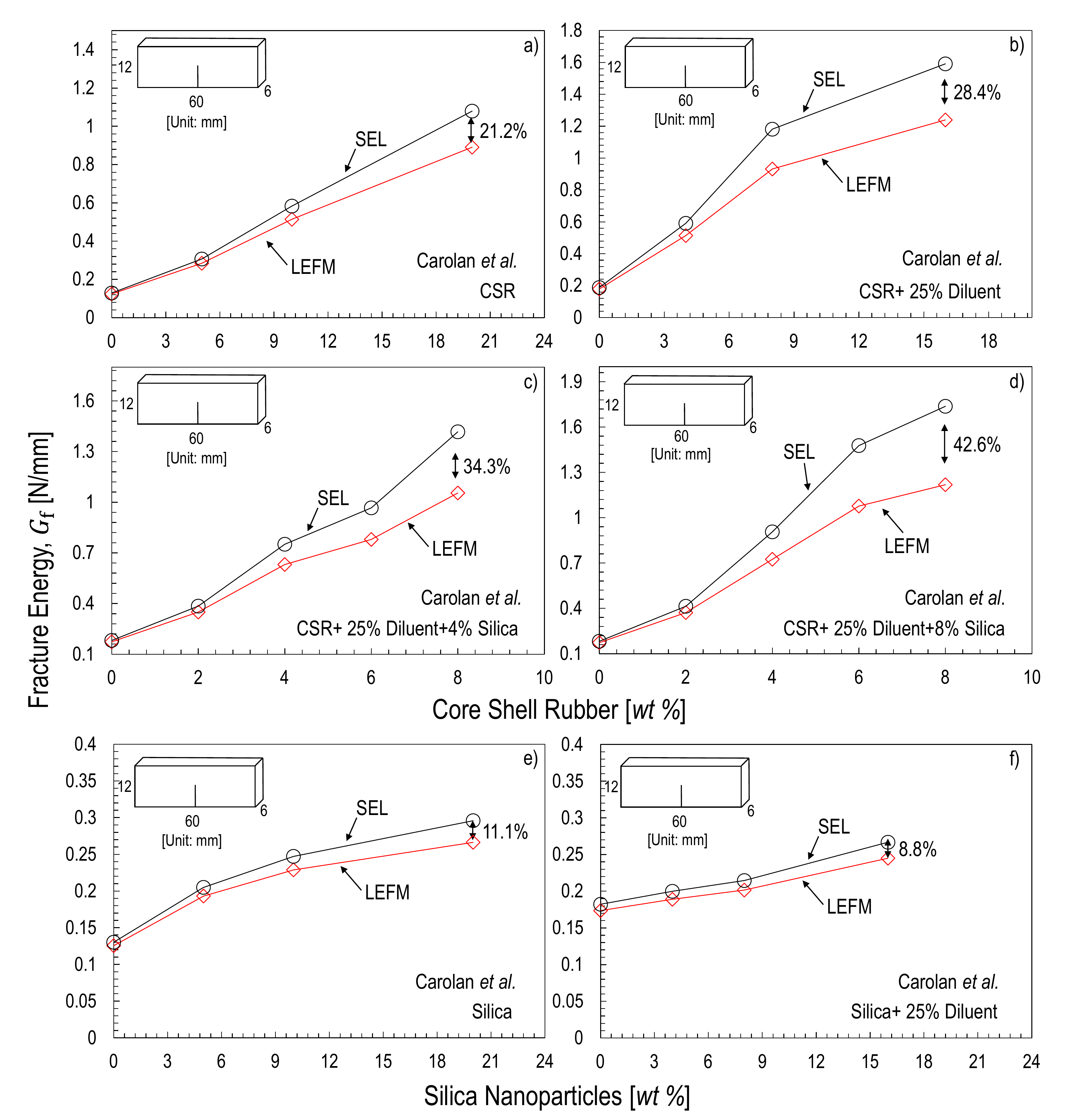} 
\caption{Mode I fracture energy estimated by Linear Elastic Fracture Mechanics (LEFM) and Size Effect Law (SEL), Eq. (\ref{eq:Gf_SEL2}). The latter formulation accounts for the finite size of the nonlinear Fracture Process Zone (FPZ) in thermoset nanocomposites. Data re-analyzed from \cite{carolan2016_1}.}
\label{fig:carolan}
\end{figure}

\begin{figure}[H]
\center
\includegraphics[scale=0.5]{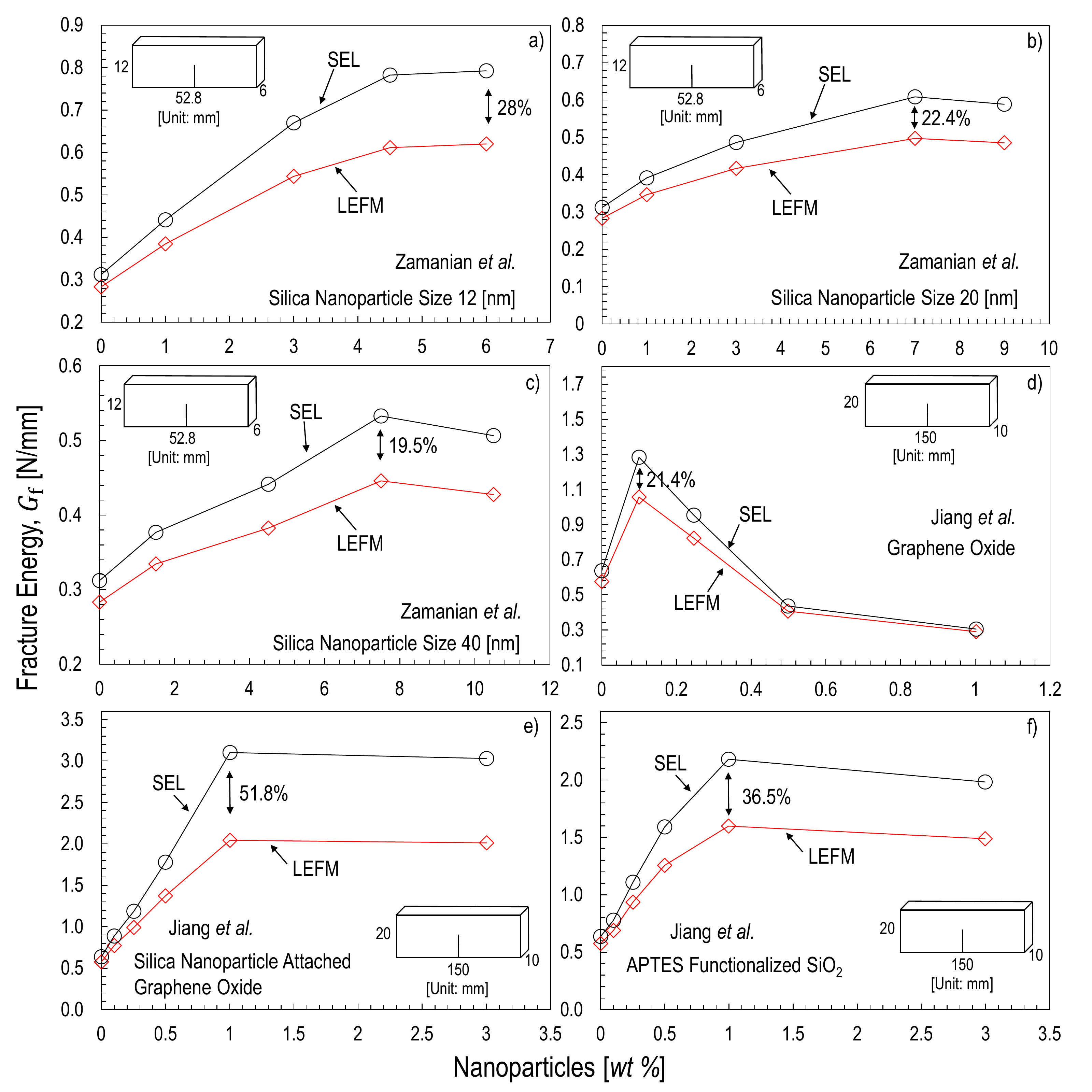}
\caption{Mode I fracture energy estimated by Linear Elastic Fracture Mechanics (LEFM) and Size Effect Law (SEL), Eq. (\ref{eq:Gf_SEL2}). The latter formulation accounts for the finite size of the nonlinear Fracture Process Zone (FPZ) in thermoset nanocomposites. Data re-analyzed from \cite{jiang2013_1} and \cite{zamanian2013_1}.}
\label{fig:zamanian&jiang}
\end{figure}

\begin{figure}[H]
\center
\includegraphics[scale=0.5]{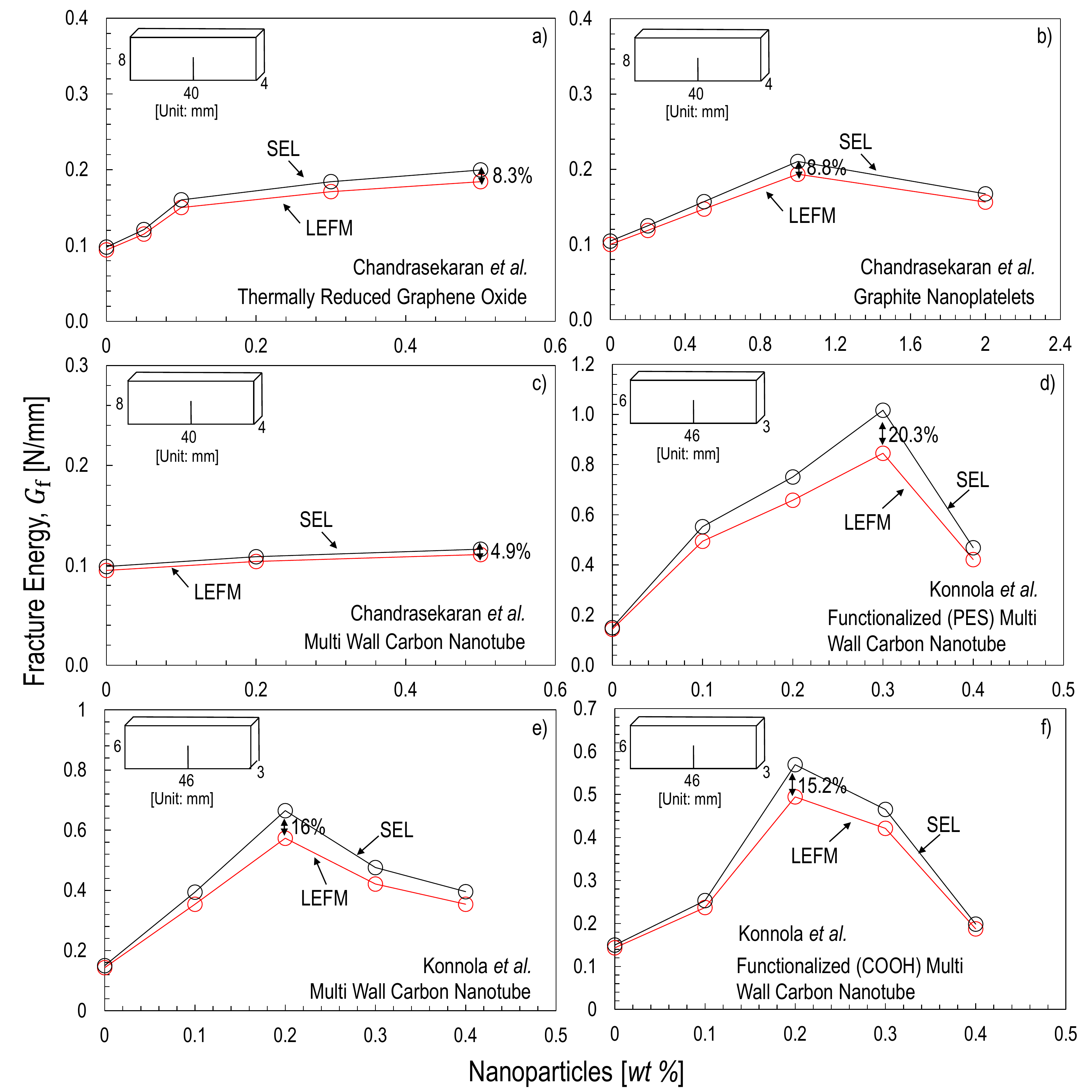}
\caption{Mode I fracture energy estimated by Linear Elastic Fracture Mechanics (LEFM) and Size Effect Law (SEL), Eq. (\ref{eq:Gf_SEL2}). The latter formulation accounts for the finite size of the nonlinear Fracture Process Zone (FPZ) in thermoset nanocomposites. Data re-analyzed from \cite{ChaSa14} and \cite{konnola2015_1}.}
\label{fig:chandra&konnola}
\end{figure}

\begin{figure}[H]
\center
\includegraphics[scale=0.5]{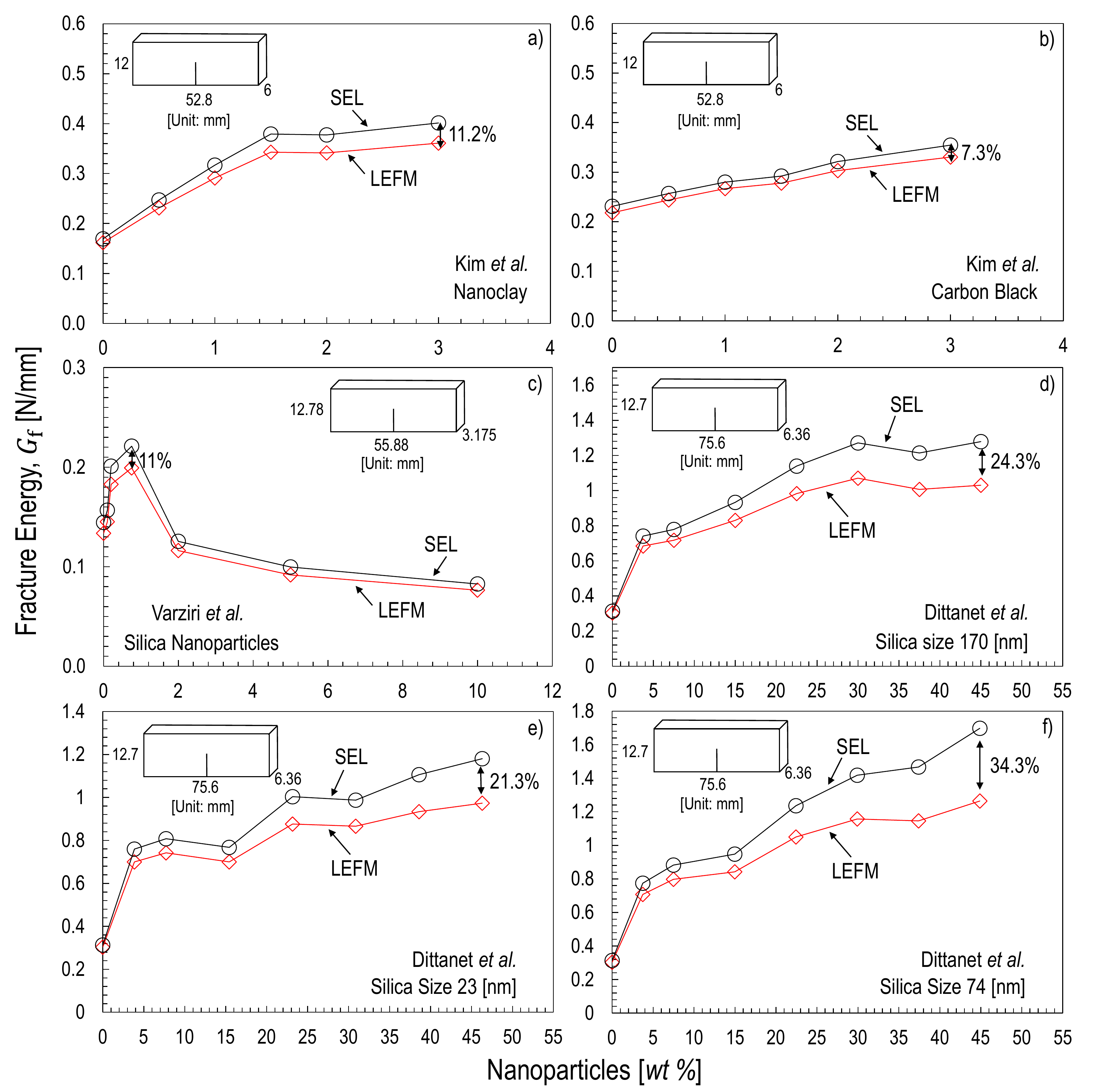}
\caption{Mode I fracture energy estimated by Linear Elastic Fracture Mechanics (LEFM) and Size Effect Law (SEL), Eq. (\ref{eq:Gf_SEL2}). The latter formulation accounts for the finite size of the nonlinear Fracture Process Zone (FPZ) in thermoset nanocomposites. Data re-analyzed from \cite{kim2008_1}, \cite{vaziri2011_1} and \cite{dittanet2012_1}.}
\label{fig:dittanet&vaziri&kim}
\end{figure}

\begin{figure}[H]
\center
\includegraphics[scale=0.5]{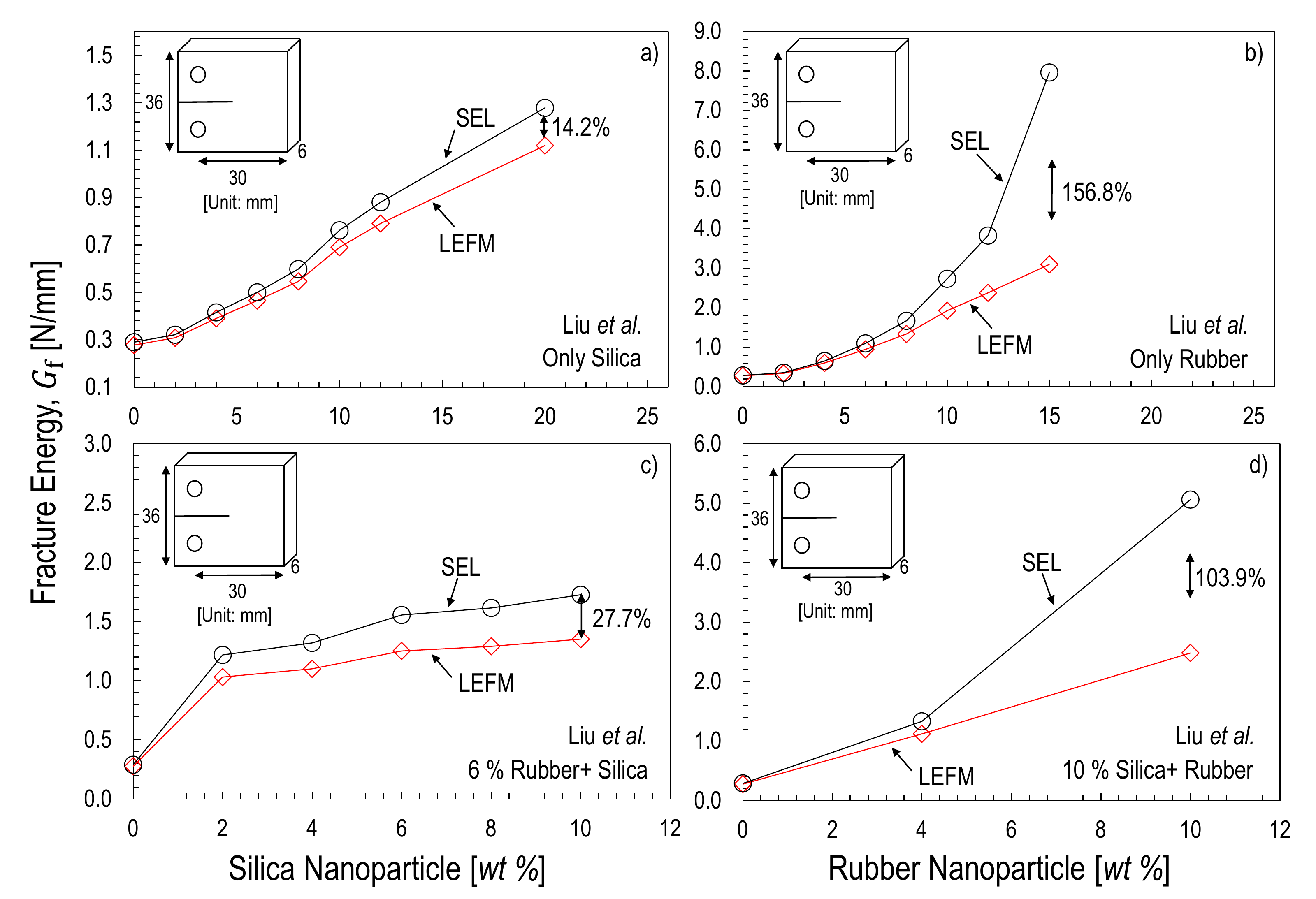}
\caption{Mode I fracture energy estimated by Linear Elastic Fracture Mechanics (LEFM) and Size Effect Law (SEL), Eq. (\ref{eq:Gf_SEL2}). The latter formulation accounts for the finite size of the nonlinear Fracture Process Zone (FPZ) in thermoset nanocomposites. Data re-analyzed from \cite{liu2011_1}.}
\label{fig:liu}
\end{figure}

Several types of nanofillers are investigated in this re-analysis including carbon-based nano-fillers (such as carbon black, graphene oxide, graphene nanoplatelets, and multi-wall carbon nanotubes), rubber and silica nanoparticles, and nanoclay. The fracture energy estimated from LEFM compared to the calculation through SEL, Eq. (\ref{eq:Gf_SEL2}), for nanomodified SENB and CT specimens are plotted in Figures \ref{fig:carolan}-\ref{fig:liu} along with the highest difference.

Figure \ref{fig:carolan} shows data elaborated from Carolan \emph{et al.} \cite{carolan2016_1} who conducted fracture tests on SENB specimens nano-modified by six different combinations of nanofillers. As can be noted, while for the pristine polymer the difference between LEFM and SEL is negligible, this is not the case for the nanomodified polymers, the difference increasing with increasing nanofiller content. The difference varies based on the type of nanofiller used, with the greatest value being 42.6\% for the addition of 8 wt\% core shell rubber mixed with 25\% diluent and 8\% silica. This confirms that for the SENB specimens tested in \cite{carolan2016_1} the nonlinear behavior of the FPZ is not negligible, leading to a more ductile behavior compared to the pristine polymer. 

Similar conclusions can be drawn based on Figures \ref{fig:zamanian&jiang}a-f which report the analysis of fracture tests conducted by Zamanian \emph{et al.} \cite{zamanian2013_1} and Jiang \emph{et al.} \cite{jiang2013_1} on polymers reinforced by silica nanoparticles and silica nanoparticle$+$graphene oxide respectively. For the data in \cite{zamanian2013_1}, the greatest percent difference of fracture energy between LEFM and SEL decreased as the size of silica nanoparticle increased, with the greatest difference being 28\% for the addition of 6 wt\% 12 nm silica nanoparticles. For all the systems investigated, the maximum deviation from LEFM is for the largest amount of nanofiller, confirming that nanomodification lead to larger FPZ sizes and more pronounced ductility. On the other hand, the data by Jiang \emph{et al.} \cite{jiang2013_1} exhibit an even larger effect of the FPZ with the greatest difference in fracture energy between LEFM and SEL reaching up to 51.8\% for silica nanoparticle attached to graphene oxide. 

A milder effect of the FPZ can be inferred from the data by Chandrasekaran \emph{et al.} \cite{ChaSa14} who investigated three types of carbon-based nano-fillers (Figure \ref{fig:chandra&konnola}): (1) thermally reduced graphene oxide; (2) graphene nanoplatelets; and (3) multi-wall carbon nanotubes. In these cases, the difference between SEL and LEFM ranges from 4.9\% to 8.8\%, the lowest difference among all the data analyzed in this study. For these systems, the specimen size compared to the size of the nonlinear FPZ is large enough to justify the use of LEFM which provided accurate and objective results. On the other hand, a more significant effect of the FPZ can be inferred from the data reported by Konnola \emph{et al.} \cite{konnola2015_1} who studied three different types of functionalized and nonfunctionalized nano-fillers. In this case, the greatest difference in fracture energy ranges between 15.2\% to 20.3\%.

SENB specimens nano-modified by nanoclay and carbon black respectively were tested by Kim \emph{et al.} \cite{kim2008_1}. As Figure \ref{fig:dittanet&vaziri&kim} shows, in this case, the specimen size is enough to justify the use of LEFM as confirmed by the low difference with SEL (11.2\% for nanoclay and 7.3\% for carbon black). Similar conclusions can be drawn on the silica nanoparticles investigated by Vaziri \emph{et al.} \cite{vaziri2011_1}. However, for the three different sizes of silica nanoparticles investigated by Dittanet \emph{et al.} \cite{dittanet2012_1}, a significant difference between LEFM and SEL is observed, confirming that these specimens belonged to the transition zone between ductile and brittle behavior where the effects of the nonlinear FPZ cannot be neglected.

Figure \ref{fig:liu} shows a re-analysis of the data reported by Liu \emph{et al.} \cite{liu2011_1} who tested CT specimens nano-modified by four different combinations of silica nanoparticle and rubber. As can be noted, in this case, the FPZ indeed affects the fracturing behavior significantly. Adopting LEFM, which assumes the size of the FPZ to be negligible, for the estimation of $G_f$ would lead to an underestimation of up to 156.8\% for the case of polymer reinforced by 15 wt\% rubber only. This tremendous difference, the largest found in the present study, gives a tangible idea on the importance of accounting for the nonlinear damage phenomena occurring in nanocomposites which can lead to a significant deviation from the typical brittle behavior of thermoset polymers.  

\section{Cohesive Zone Modeling of Thermoset Nanocomposites}

\begin{figure}[H]
\center
\includegraphics[scale=0.6]{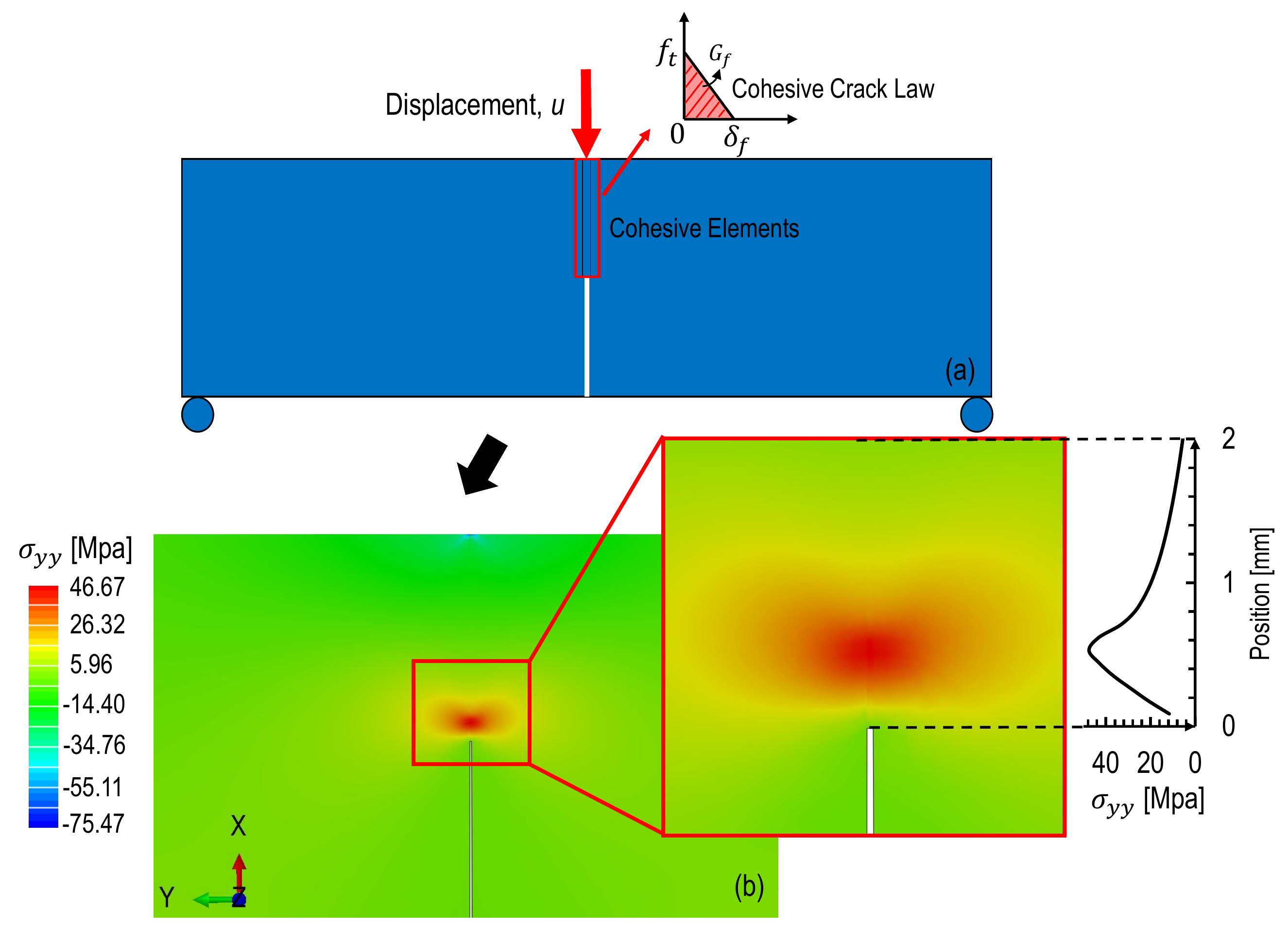}
\caption{(a) Schematic geometry with cohesive crack modeling; (b) Cohesive crack modeling and stress profile in front of crack tip.}
\label{fig:cohesive}
\end{figure}

To have a deeper understanding of the fracturing behavior of nanocomposites, a computational investigation is conducted leveraging a cohesive zone model featuring a Linear Cohesive Law (LCL) in ABAQUS Explicit 2017. To this end, as illustrated in Figure \ref{fig:cohesive}, a Single Edge Notch Bending (SENB) specimen is simulated by 4-node two-dimensional cohesive elements (COH2D4) with a traction-separation law to model the crack and 4-node bi-linear plain strain quadrilateral elements (CPE4R) with a linear elastic isotropic behavior to model the rest of the specimen. The width of crack is modeled as 4 $\mu$m based on the image obtained from Scanning Electron Microscopy (SEM).

\subsection{Linear Cohesive Crack Law} 
\label{sec:LCCL}
\begin{figure}[H]
\center
\includegraphics[scale=0.7]{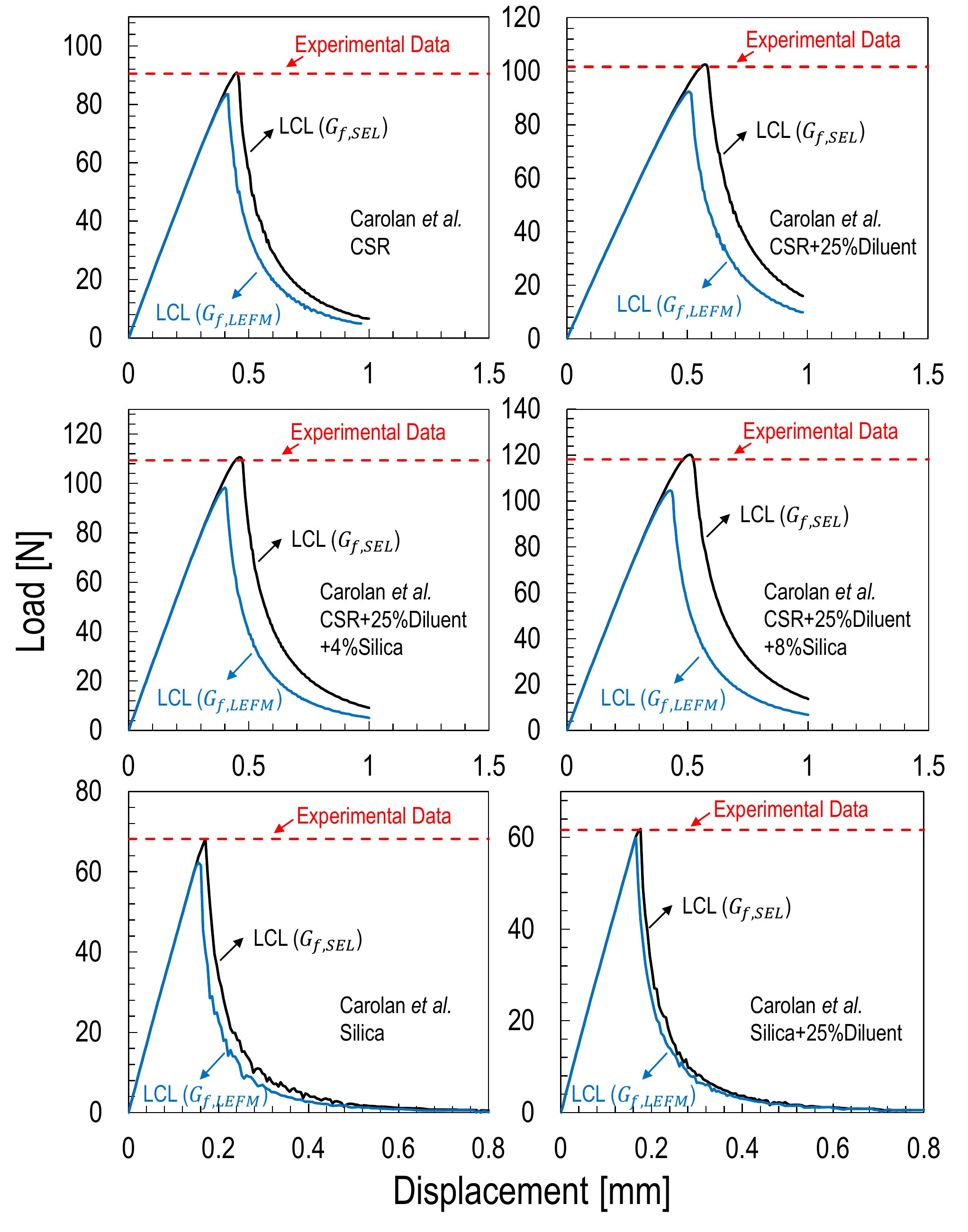}
\caption{Load-displacement curves obtained by using a linear cohesive crack law with both LEFM and corrected $G_{f}$ on the re-analysis of data from \cite{carolan2016_1}.}
\label{fig:loaddiscarolan}
\end{figure}

\begin{figure}[H]
\center
\includegraphics[scale=0.7]{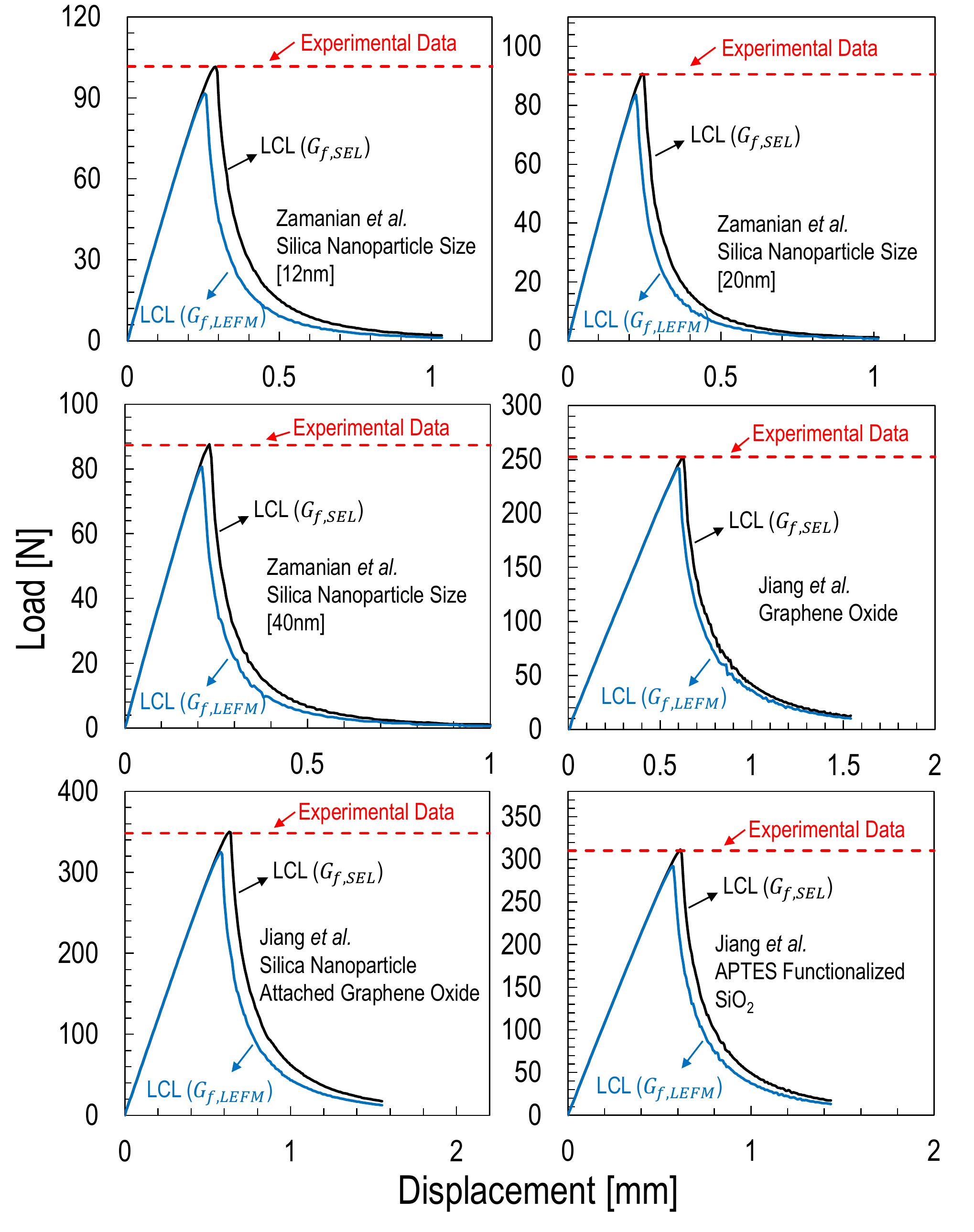}
\caption{Load-displacement curves obtained by using a linear cohesive crack law with both LEFM and corrected $G_{f}$ on the re-analysis of data from \cite{jiang2013_1}, \cite{zamanian2013_1}.}
\label{fig:loaddiszemanjiang}
\end{figure}

\begin{figure}[H]
\center
\includegraphics[scale=0.7]{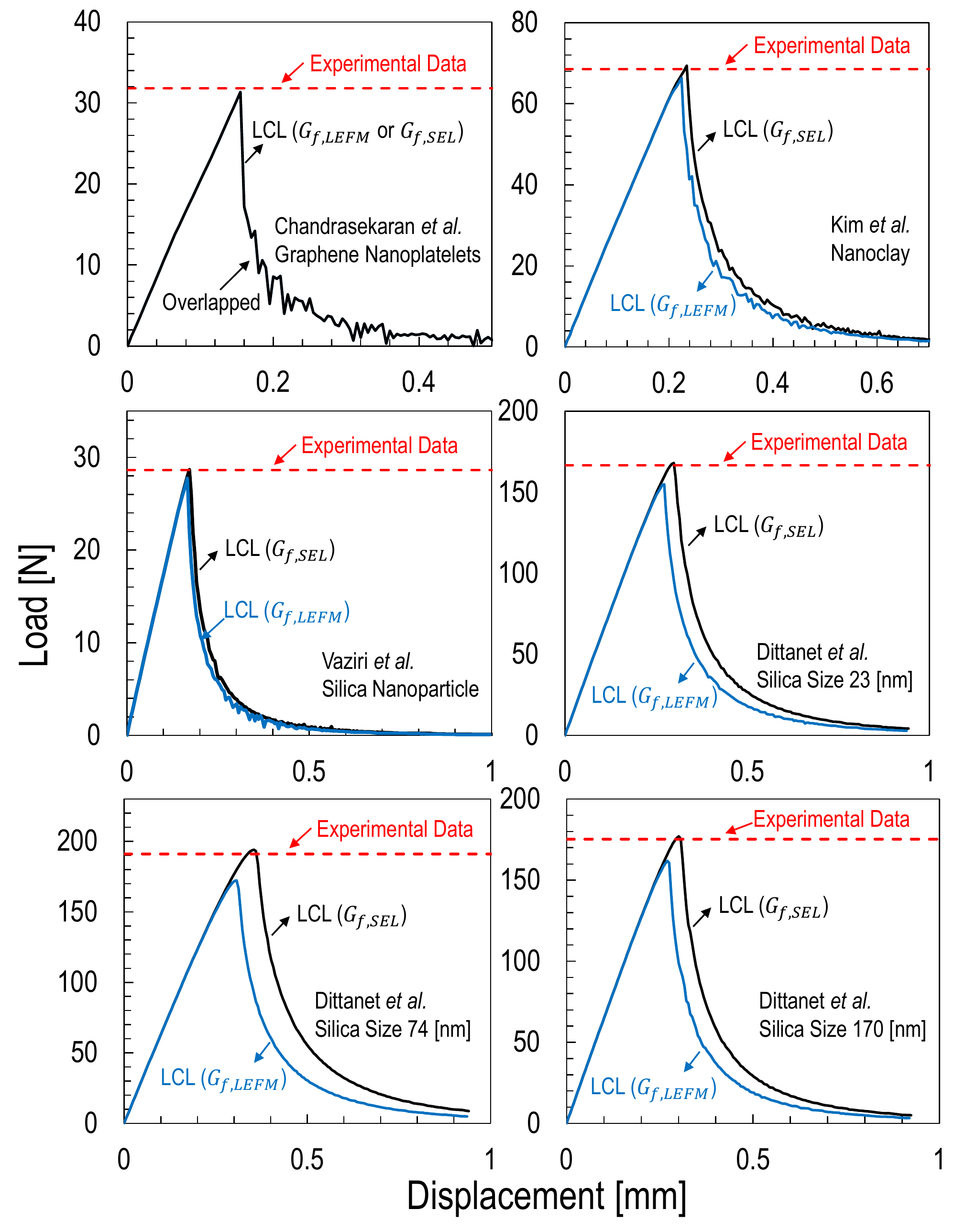}
\caption{Load-displacement curves obtained by using a linear cohesive crack law with both LEFM and corrected $G_{f}$ on the re-analysis of data from \cite{ChaSa14,kim2008_1,vaziri2011_1,dittanet2012_1}.}
\label{fig:loaddischankimvadi}
\end{figure}

\begin{figure}[H]
\center
\includegraphics[scale=0.7]{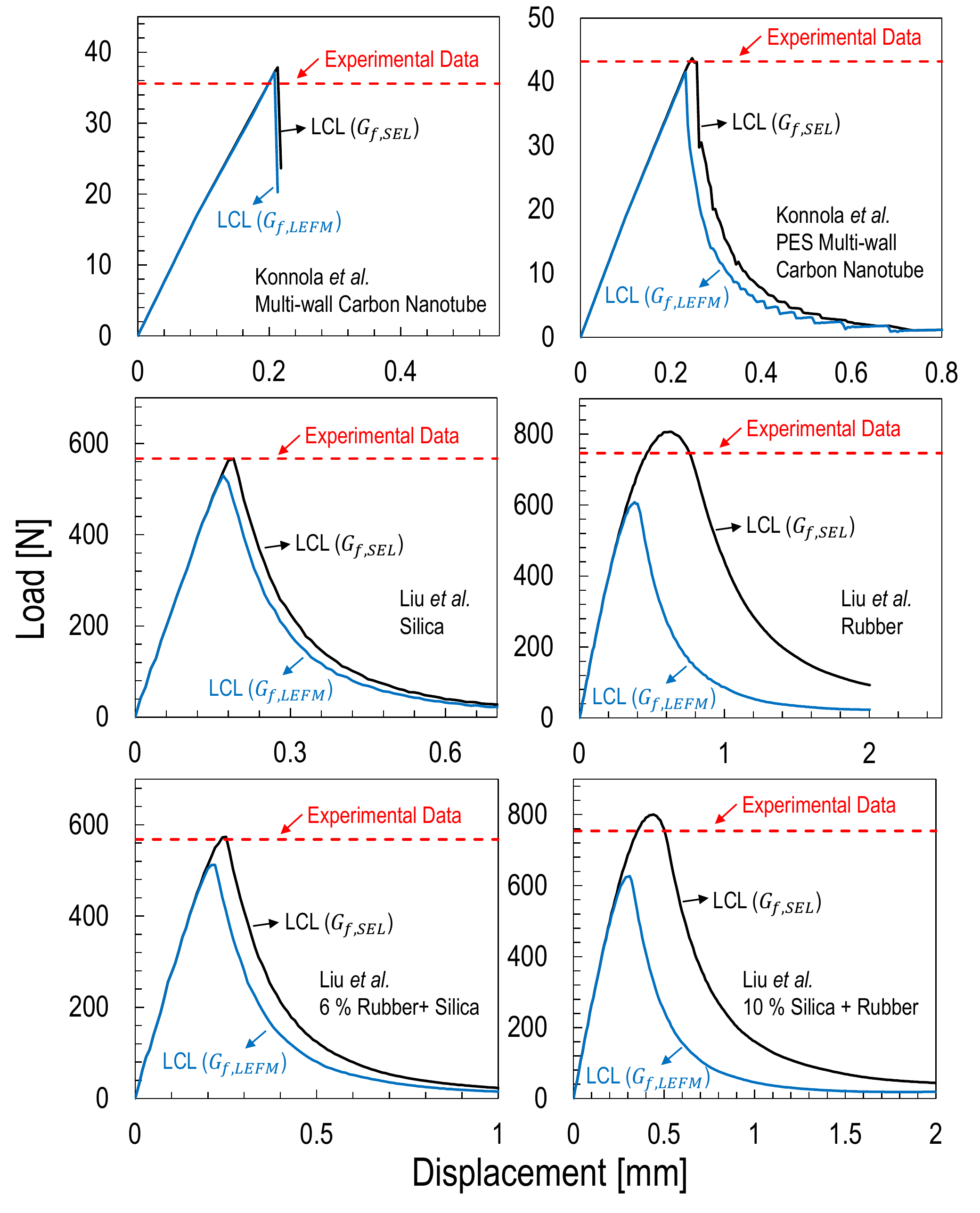}
\caption{Load-displacement curves obtained by using a linear cohesive crack law with both LEFM and corrected $G_{f}$ on the re-analysis of data from \cite{konnola2015_1}, \cite{liu2011_1}.}
\label{fig:loaddiskonnoliu}
\end{figure}

To corroborate the results discussed in the forgoing sections, the analysis by means of a cohesive zone model is carried out using the fracture energy estimated via LEFM, $G^{LEFM}_{f}$, and the one calculated through Eq.(\ref{eq:Gf_SEL2}), $G^{SEL}_{f}$. As can be noted from Figures \ref{fig:loaddiscarolan}-\ref{fig:loaddiskonnoliu}, the cohesive zone model using $G^{SEL}_{f}$ as input shows an excellent agreement with the experimental data in the literature. However, this is not the case if $G_{f}$ by LEFM is used. In fact, this is a further confirmation that Size Effect Law (SEL) can be adapted to re-analyze the fracture tests available in the literature.

\begin{figure}[H]
\center
\includegraphics[scale=0.645]{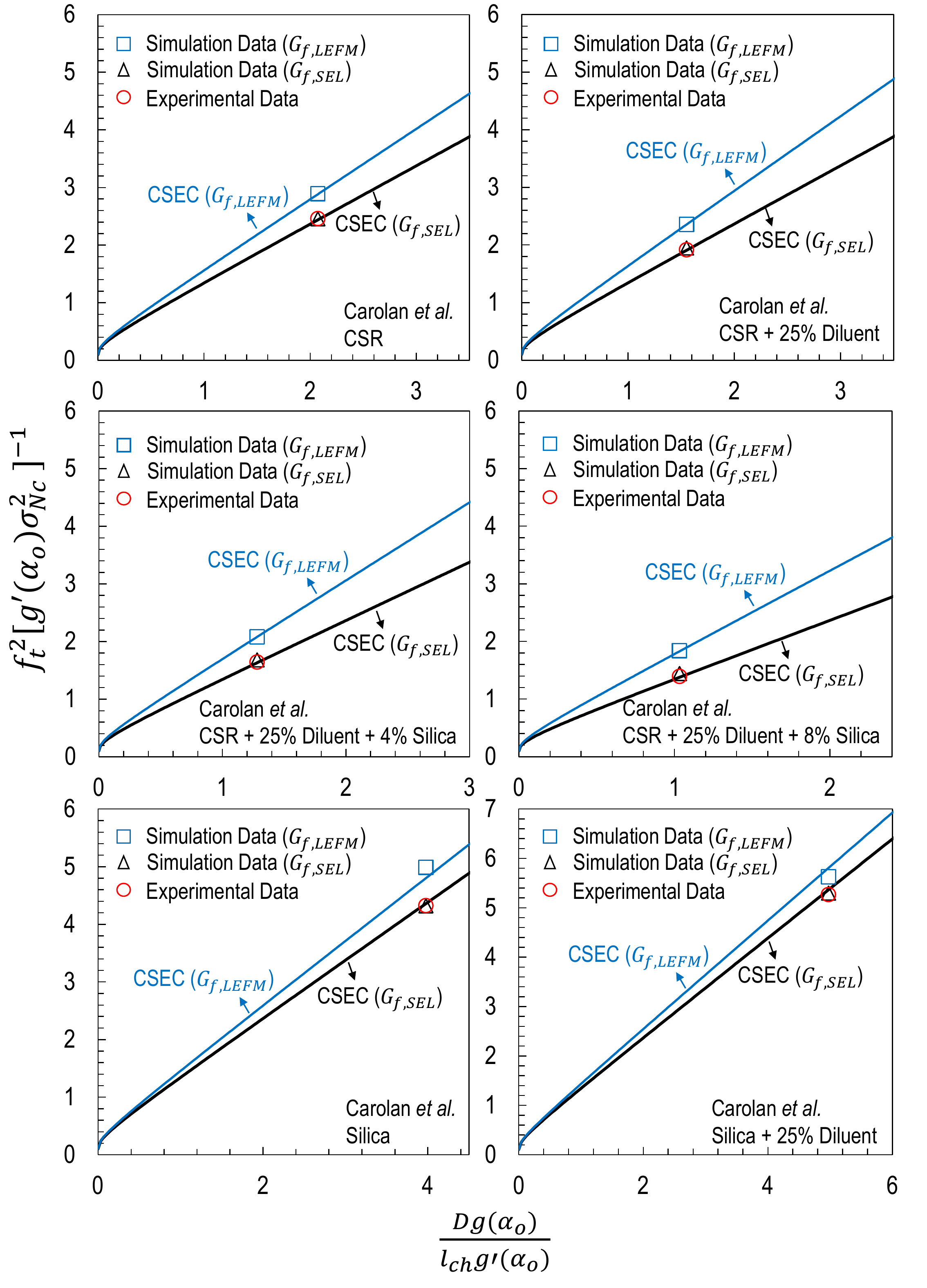}
\caption{Comparison between LCL results and experimental data from \cite{carolan2016_1}.}
\label{fig:cohesivecarolan}
\end{figure}

\begin{figure}[H]
\center
\includegraphics[scale=0.64]{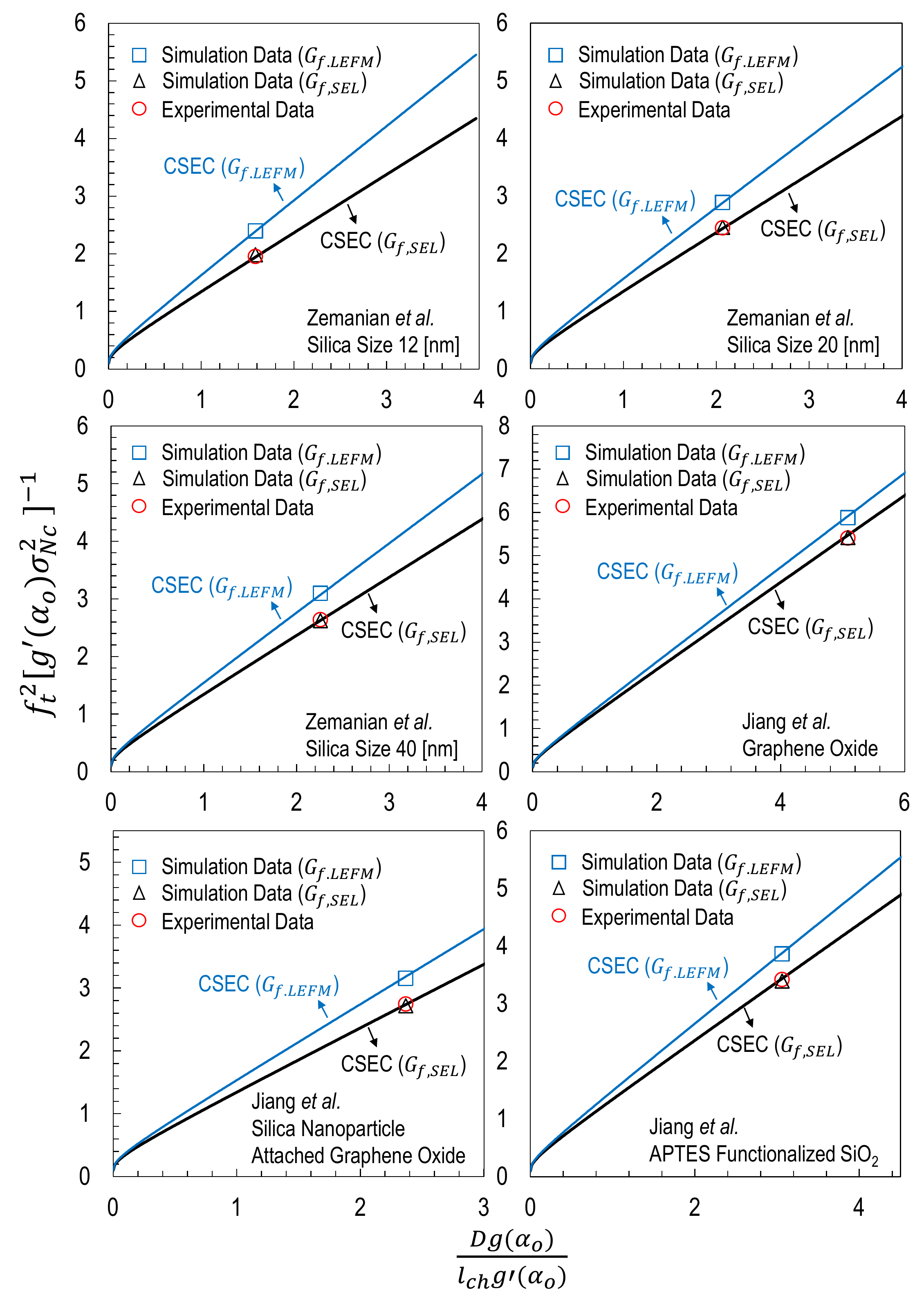}
\caption{Comparison between LCL results and experimental data from \cite{jiang2013_1}, \cite{carolan2016_1}, \cite{zamanian2013_1}.}
\label{fig:cohesivezemanjiang}
\end{figure}

\begin{figure}[H]
\center
\includegraphics[scale=0.64]{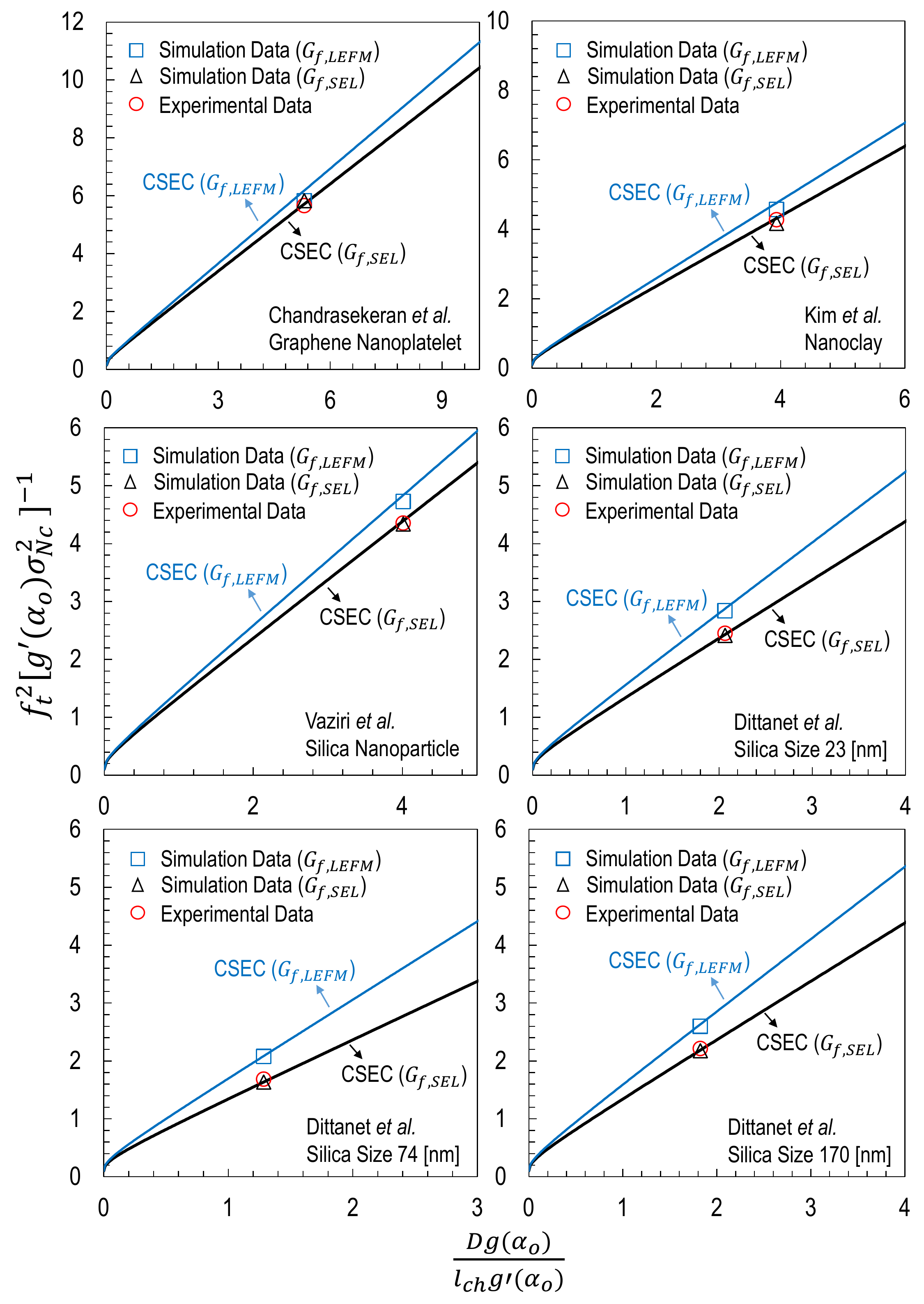}
\caption{Comparison between LCL results and experimental data from \cite{ChaSa14,kim2008_1,vaziri2011_1,dittanet2012_1}.}
\label{fig:cohesivechankimvadi}
\end{figure}

\begin{figure}[H]
\center
\includegraphics[scale=0.655]{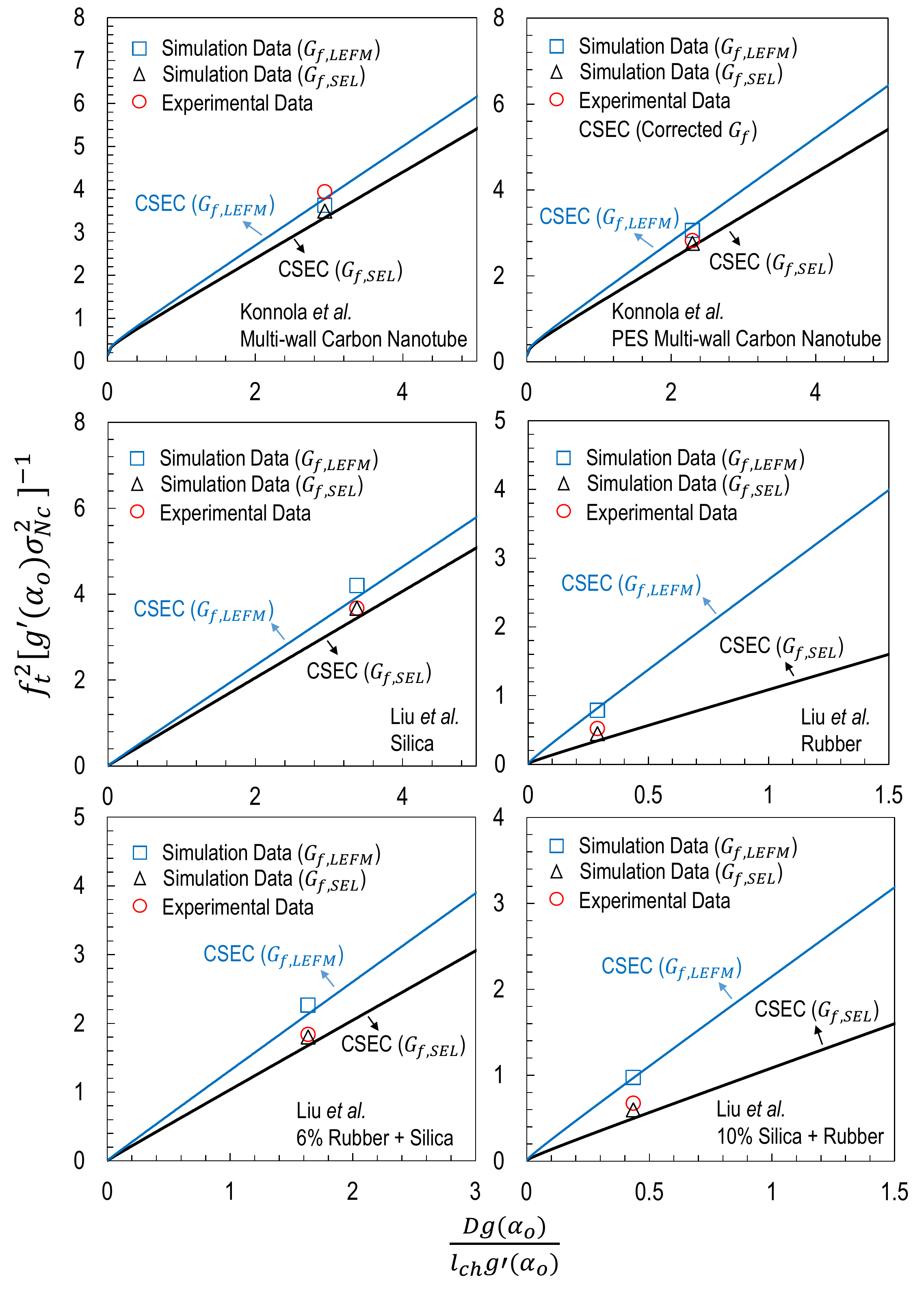}
\caption{Comparison between LCL results and experimental data from \cite{konnola2015_1}, \cite{liu2011_1}.}
\label{fig:cohesivekonnoliu}
\end{figure}

By leveraging a linear cohesive crack modeling with the corrected $G_{f}$, the fracturing behavior on the scaling of nanocomposites can be predicted without additional tests in the lab. As  Figures \ref{fig:cohesivecarolan}-\ref{fig:cohesivekonnoliu} show, experimental data in the literature and simulation results using a linear cohesive crack law are plotted along with the analytical expression for Cohesive Size Effect Curves (CSEC) purposed by Cusatis \emph{et al.} \cite{cusatis2009_1}. In these Figures, it can be noted that, for large specimen sizes, the prediction on the peak load of investigated nanocomposites by using LEFM $G_{f}$ leads to a significant underestimation. It is worth mentioning here that this analysis is on the assumption that nanocomposites in the literature follow a linear cohesive law.

\subsection{Bi-linear Cohesive Crack Law} 
\label{sec:BCCL}

\begin{figure}[H]
\center
\includegraphics[scale=0.75]{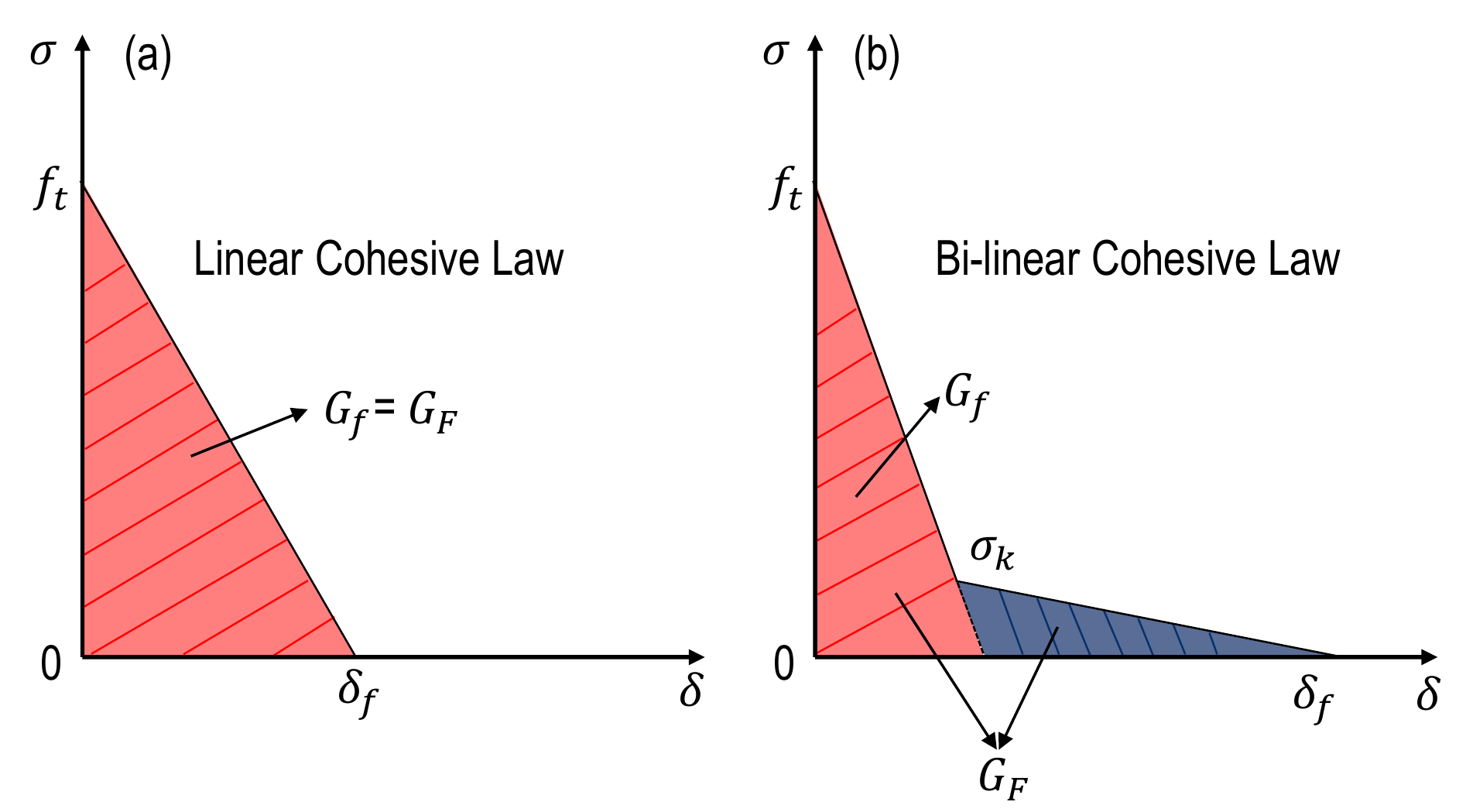}
\caption{Cohesive laws used in this study: (a) Linear Cohesive Law (LCL); (b) Bi-linear Cohesive Law (BCL).}
\label{fig:linearandbilinear}
\end{figure}

Thanks to the comprehensive investigation on the size effect in graphene nanocomposites by  Mefford \textit{et al} \cite{CoryandYao}, the characteristics of the cohesive crack law can be further studied. To this end, both linear and bi-linear cohesive laws with the same fracture energy are used to match load-displacement curves obtained from experimental fracture tests on geometrically scaled Single Edge Notch Bending (SENB) specimens with varying contents of graphene. As illustrated in Figure \ref{fig:linearandbilinear}, a linear cohesive law can be described through two parameters: (a) tensile strength, $f_t$ and (b) fracture energy, $G_f$, which represents the area under the linear cohesive law. On the other hand, a bilinear cohesive law requires four parameters: (a) tensile strength $f_t$, (b) initial fracture energy,  $G_f$, which represents the area under the initial segment of the bi-linear cohesive law; (b) total fracture energy, $G_F$, which is the total area under the bi-linear cohesive law; (d) change-of-slope stress, $\sigma_k$, which is the value of stress at the intersection of the initial and tail segments. It is worth mentioning here that, for the bi-linear cohesive law, different intersection points are investigated in order to match experimental load-displacement curves. 

\begin{figure}[H]
\center
\includegraphics[scale=0.55]{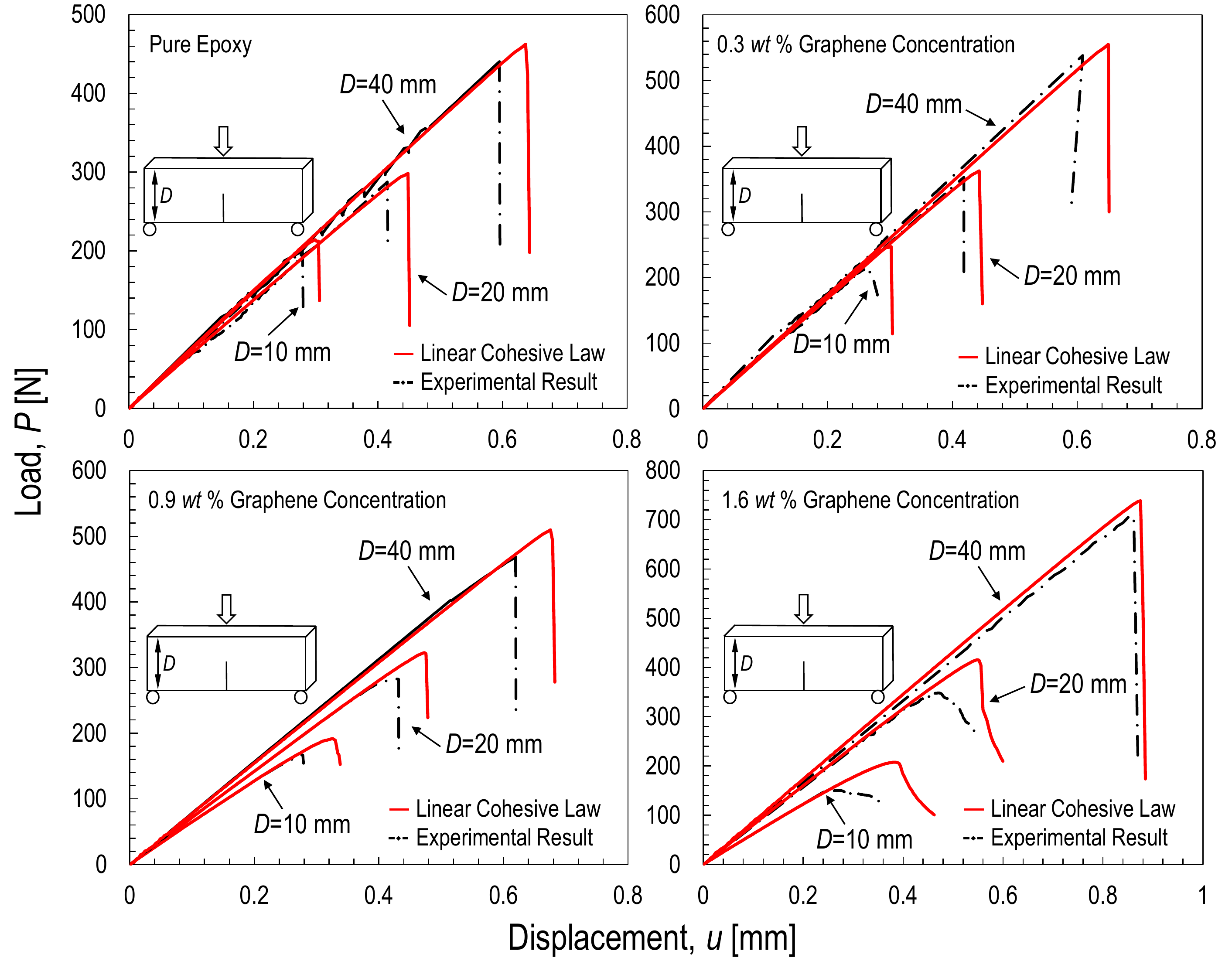}
\caption{Load-displacement curves vs. cohesive zone model featuring a Linear Cohesive Law (LCL) for different graphene contents and specimen sizes.}
\label{fig:linear}
\end{figure}

\begin{figure}[H]
\center
\includegraphics[scale=0.55]{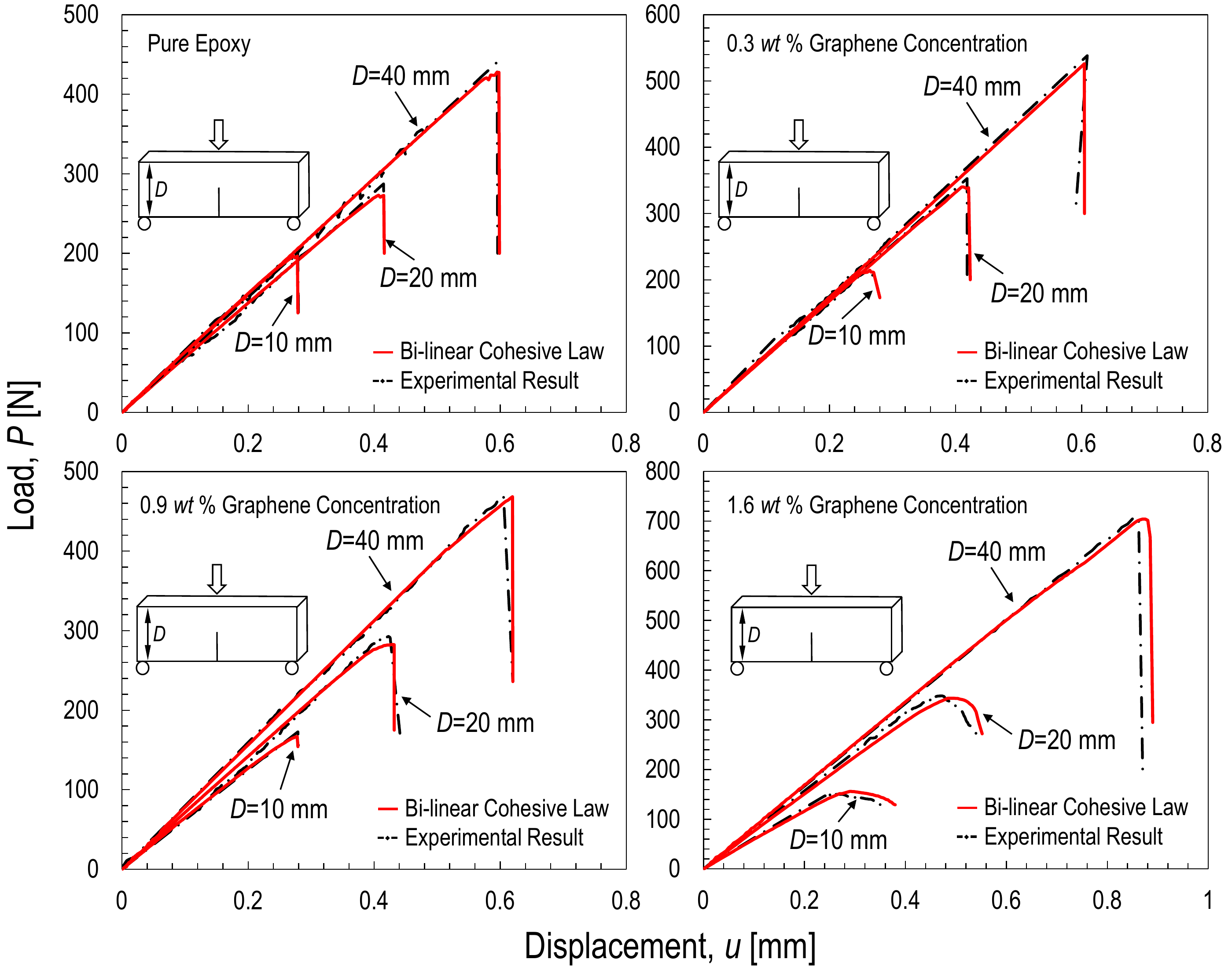}
\caption{Load-displacement curves vs. cohesive zone model featuring a Bi-linear Cohesive Law (BCL) for different graphene contents and specimen sizes.}
\label{fig:bilinear}
\end{figure}

\begin{figure}[H]
\center
\includegraphics[scale=0.55]{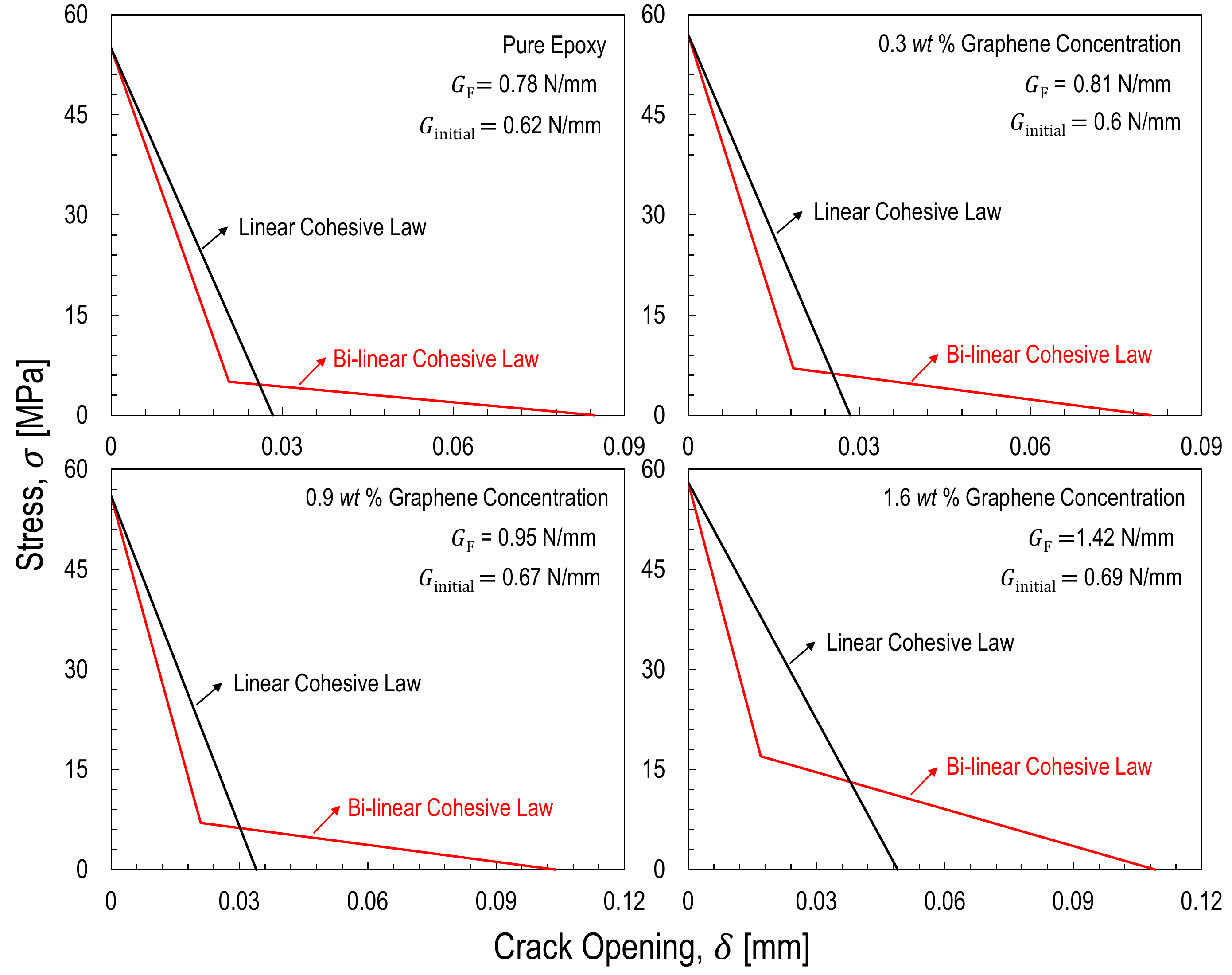}
\caption{Calibrated bi-linear cohesive law vs. linear cohesive law with the same total fracture energy.}
\label{fig:soft}
\end{figure}

Figures \ref{fig:linear}-\ref{fig:bilinear} show a comparison between the experimental load defection curves and simulation through a Cohesive Zone Model (CZM) featuring a linear and bi-linear cohesive law respectively. It can be noted that, while the bi-linear cohesive law successfully matches experimental curves of specimens with different sizes and graphene concentrations, this is not the case for the linear cohesive law, with a significant underestimation of the experimental curves. In fact, the bi-linear cohesive law provides a very accurate description of fracture tests with errors on structural strength less than $7$\% whereas the linear cohesive law shows a maximum deviation from the tests of $30$\%. This result suggests that a bi-linear cohesive law may be better suited for the description of the cohesive fracture behavior of nanocomposites, although a linear cohesive law may still provide reasonable results and can be used for a preliminary, course, approximation. Further size effect studies on different material systems will shed more light on this important aspect. A comparison between the calibrated linear and bi-linear cohesive laws for the various graphene contents is shown in Figures \ref{fig:soft}a-d.  

\begin{figure}[H]
\center
\includegraphics[scale=0.95]{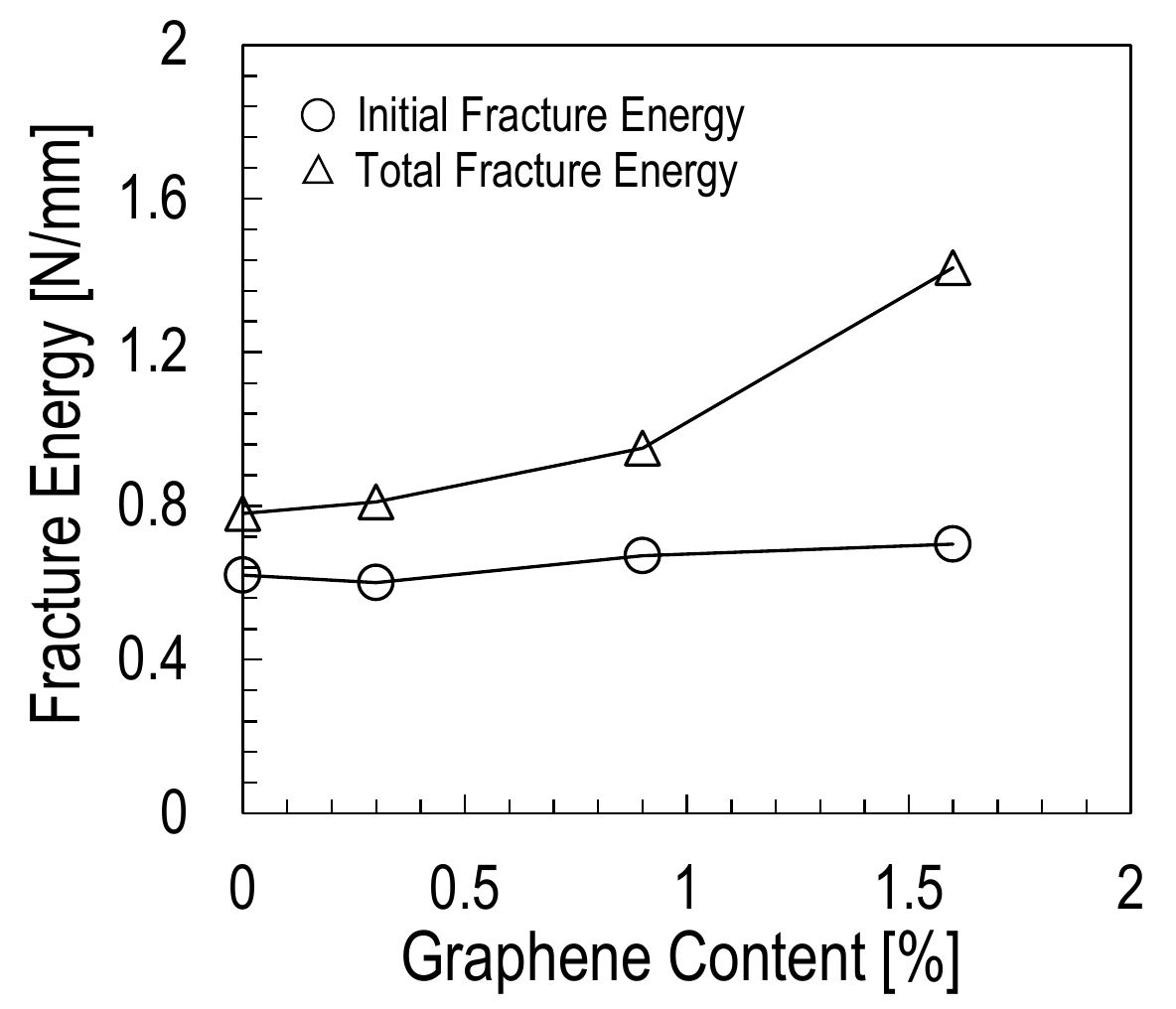}
\caption{Comparison between initial and total fracture energy of the calibrated bi-linear cohesive law.}
\label{fig:initial}
\end{figure}

\begin{figure}[H]
\center
\includegraphics[scale=0.7]{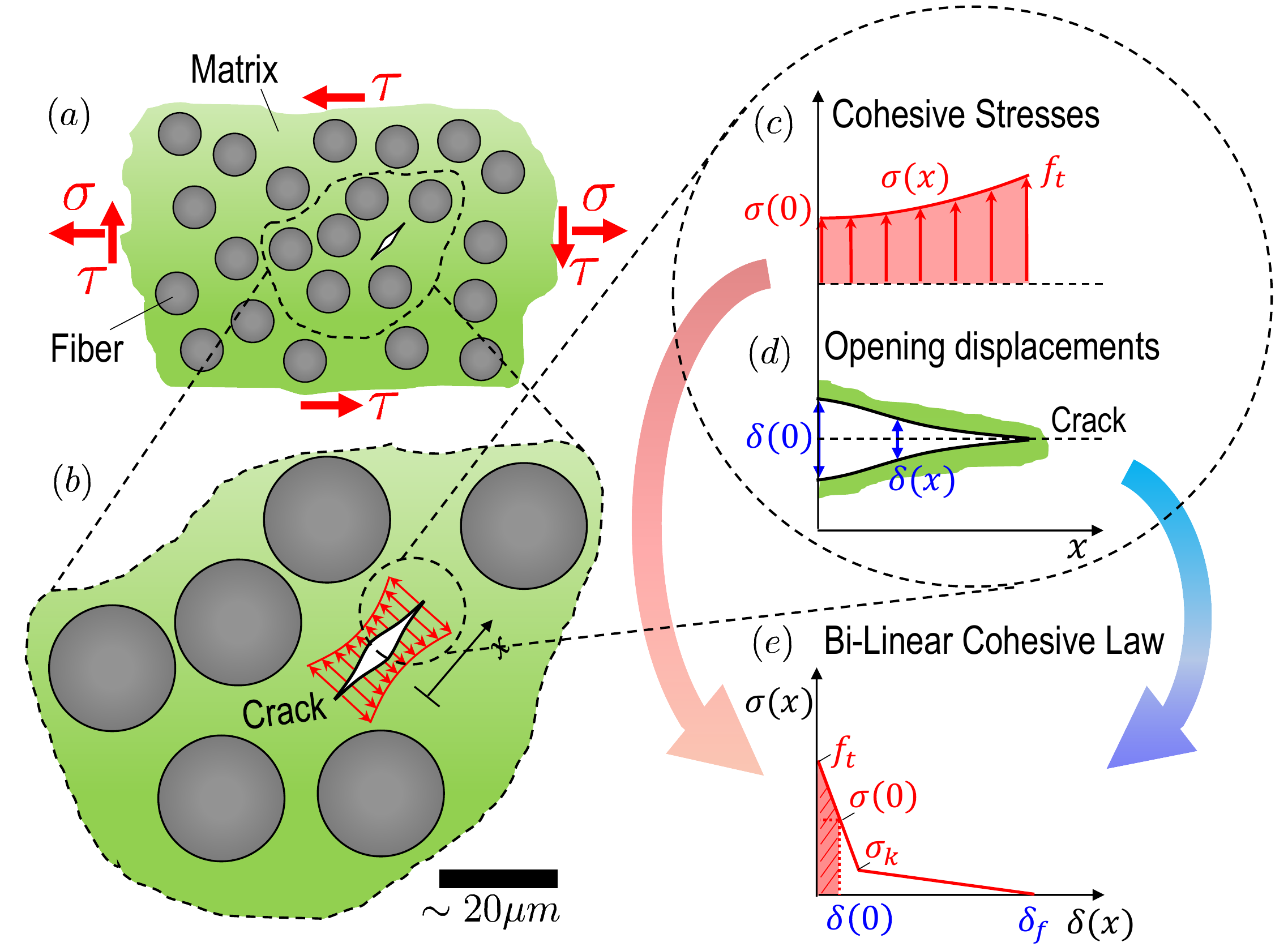}
\caption{Schematic representation of a micro-crack propagating in a composite: (a) and (b) Cohesive crack formation; (c) cohesive stresses bridging the crack faces; (d) distribution of crack opening displacement and (e) corresponding stresses and displacements in the cohesive law. Note that, for a micro-crack, the cohesive stresses do not enter the second arm of the bi-linear cohesive law ($\sigma_{min}>\sigma_{k}$)}
\label{fig:micro}
\end{figure}

Figure \ref{fig:initial} shows the initial, $G_f$, and total, $G_F$, fracture energy as a function of graphene platelet content. It is interesting to note that the initial fracture energy does not increase significantly as a function of graphene content. The increasing total fracture energy for higher graphene contents can all be ascribed to the change in slope of the second part of the curve. This is the indication that, for crack opening displacements lower than about $20 \mu$m, which is e.g. the case of a crack propagating between micrometer fibers in a unidirectional composite (Figure \ref{fig:micro}), the effects of nanomodification may be negligible. In fact, as schematically explained in inserts (b) and (c) of Figure \ref{fig:micro}), in such a case only the initial part of the bi-linear cohesive law is developed and drives the fracturing behavior. This may explain why the use of nanomodification to improve the fracturing behavior of the polymer matrix in fiber composites has met with changing fortunes. Even if increments of the total fracture energy by nanomodification can be observed from tests on laboratory-scale specimens, this does not guarantee that the initial portion of the cohesive curve, which drives the microcracking in composites, has improved. This latter aspect can be clarified only by size effect testing and cohesive zone modeling, as clearly shown in this work.

\section{Conclusions}
Leveraging a large bulk of literature data, this paper investigated the effects of the Fracture Process Zone (FPZ) on the fracturing behavior of thermoset polymer nanocomposites, an aspect of utmost importance for structural design but so far overlooked. Based on the results obtained in this study, the following conclusions can be elaborated:

1. The double logarithmic plots of the normalized structural strength as a function of the normalized characteristic size of geometrically-scaled SENB specimens show that the experimental data on nanocomposites available in the literature are in excellent agreement with Size Effect Law (SEL). Most of nanocomposites are located in the transitional range in which the fracturing behavior cannot be characterized by Linear Elastic Fracture Mechanics (LEFM);

2. Size Effect Law and Cohesive Zone Modeling show that for most of the fracture tests on polymer nanocomposites investigated in this work, the effects of the nonlinear FPZ are not negligible, leading to significant deviations from LEFM. As the data indicate, this aspect needs to be taken into serious consideration since the use of LEFM to estimate mode I fracture energy can lead to an error as high as $156$\% depending on the specimen size and nanofiller content;

3. The deviation from LEFM reported in the re-analyzed results is related to the size of the Fracture Process Zone (FPZ) for increasing contents of nanofiller. In the pristine polymer the damage/fracture zone close to the crack tip, characterized by significant non-linearity due to subcritical damaging, is generally very small compared to the specimen sizes investigated. This is in agreement with the inherent assumption of LEFM of negligible non-linear effects during the fracturing process. However, the addition of nano-fillers results in larger and larger FPZs. For sufficiently small specimens, the size of the highly non-linear FPZ is not negligible compared to the specimen characteristic size thus highly affecting the fracturing behavior, this resulting into a significant deviation from LEFM;
 
4. To get a deeper understanding of the cohesive behavior of nanocomposites, recent fracture tests on thermosets reinforced by graphene nanoplatelets were re-analyzed via a cohesive model featuring a bi-linear law for all the sizes and graphene contents considered. It is concluded that, in general, a bi-linear cohesive law provides a very accurate description of fracture tests with errors on structural strength less than $7$\%. A reasonable agreement is also found leveraging a linear cohesive law with errors on structural strength no larger than $30$\%.

5. The analysis via the bi-linear cohesive law provided unprecedented insights on the influence of graphene nanoplatelets on the cohesive stresses. It is found that, for the size range investigated, the initial part of the cohesive law is unaffected by nanomodification. The increasing total fracture energy for higher graphene contents can all be ascribed to the change in slope of the second part of the cohesive law. Two main considerations can be made from this result: (a) the toughening by graphene nanoplatelets requires sufficiently large crack opening displacement (larger than about $20\mu$m for the system investigated in this work), confirming that mechanisms such as crack deflection and splitting are the main sources of energy dissipation; (b) for very small crack opening displacements, such as for the case of a crack propagating between micrometer fibers in a composite, the effect of nanomodification may be negligible, since no change in the cohesive behavior is induced by graphene nanoplatelets in that regime. Of course, different nanoparticles and manufacturing processes may affect the initial portion of the cohesive law differently. Future work will focus on understanding the physical relation between the characteristics of the cohesive law and the nano/microstructure of the material.





\section*{Acknowledgments}
Marco Salviato acknowledges the financial support from the Haythornthwaite Foundation through the ASME Haythornthwaite Young Investigator Award and from the University of Washington Royalty Research Fund. This work was also partially supported by the William E. Boeing Department of Aeronautics and Astronautics as well as the College of Engineering at the University of Washington through Salviato's start up package.


\newpage
\section*{Figures}

\newpage

\newpage

\newpage

\newpage

\newpage

\newpage

\newpage

\newpage

\newpage

\newpage

\newpage

\newpage

\newpage

\newpage

\newpage

\newpage

\newpage

\newpage

\newpage

\newpage

\newpage


\end{document}